\documentclass[fleqn,usenatbib]{mnras}

\usepackage{subfig}
\usepackage{graphicx}
\usepackage{amsfonts}
\usepackage{lastpage}
\usepackage{newtxtext,newtxmath}
\usepackage{hyperref}
\graphicspath{{./}{./figures/}}

\usepackage[T1]{fontenc}
\usepackage{ae,aecompl}

\title[Optimising LSST Observing Strategy: WL Systematics]{Optimising LSST Observing Strategy for Weak Lensing Systematics}

\author[]{Husni Almoubayyed$^1$\thanks{\tt halmouba@andrew.cmu.edu},
Rachel Mandelbaum$^1$,
Humna Awan$^2$, Eric Gawiser$^2$, \newauthor R. Lynne Jones$^3$,
Joshua Meyers$^4$,
J. Anthony Tyson$^5$, Peter Yoachim$^6$,\newauthor The LSST Dark Energy Science Collaboration. \\
$^1$McWilliams Center for Cosmology, Department of Physics, Carnegie Mellon University, 5000 Forbes Ave, Pittsburgh, PA 15213.\\
$^2$Department of Physics and Astronomy, Rutgers University, Piscataway, NJ 08854.\\
$^3$Department of Astronomy and DIRAC Institute, University of Washington, Seattle, WA 98115. \\
$^4$Lawrence Livermore National Lab, 7000 East Ave, Livermore, CA 94550. \\
$^5$Physics Department, University of California, Davis, CA 95616. \\
$^6$Department of Astronomy, University of Washington, Seattle, WA 98115.
}

\date{\today}

\begin{document}

\label{firstpage}
\pagerange{\pageref{firstpage}--\pageref{LastPage}}
\maketitle

\begin{abstract}
The LSST survey will provide  unprecedented statistical power for measurements of dark energy. Consequently, controlling systematic uncertainties is becoming more important than ever.  The LSST observing strategy will affect the statistical uncertainty and systematics control for many science cases; here, we focus on weak lensing systematics.
The fact that the LSST observing strategy involves hundreds of visits to the same sky area provides new opportunities for systematics mitigation. We explore these opportunities by testing how different dithering strategies (pointing offsets and rotational angle of the camera in different exposures) affect additive weak lensing shear systematics on a baseline operational simulation, using the $\rho-$statistics formalism. Some dithering strategies improve systematics control at the end of the survey by a factor of up to $\sim 3-4$ better than others.  We find that a random translational dithering strategy, applied with random rotational dithering at every filter change, is the most effective of those strategies tested in this work at averaging down systematics. Adopting this dithering algorithm, we explore the effect of varying the area of the survey footprint, exposure time, number of exposures in a visit, and exposure to the Galactic plane. We find that any change that increases the average number of exposures (in filters relevant to weak lensing) reduces the additive shear systematics. Some ways to achieve this increase may not be favorable for the weak lensing statistical constraining power or for other probes, and we explore the relative trade-offs between these options given constraints on the overall survey parameters. 
\end{abstract}


\section{Introduction}

Gravitational lensing, the deflection of light paths due to the presence of a nearby mass, or weak lensing (WL) in the weak regime, has become one of the most sensitive probes of cosmological parameters \citep{Weinberg}. 
In contrast to strong lensing, WL is a statistical effect measured from very small but coherent effects on a large number of galaxies. Therefore, future large surveys of galaxies provide an opportunity for significant improvement: the report of the Dark Energy Task Force\footnote{\url{https://arxiv.org/abs/astro-ph/0609591}} shows that Stage IV experiments, such as the Vera C. Rubin Observatory Legacy Survey of Space and Time (LSST), will provide an improvement of 5--8 times over Stage II surveys with respect to the Dark Energy Task Force figure of merit (FoM).  The FoM is the reciprocal of the area enclosing the 95\% confidence set contours in the $w_0, w_a$ plane, where $w_0$ is the present value of the dark energy equation of state parameter, and $w_a$ determines its dependence on scale factor, defining the dark energy equation of the state $w = w_0 + w_a (1+a)$.
This improvement comes from breaking degeneracies using multiple dark energy probes, and is larger than the improvement in constraining power from each cosmological probe individually
\citep[e.g.,][]{2018RPPh...81f6901Z}.

Given the sizeable statistical power that the LSST
provides \citep{ScienceBook,DESCSRD,Overview}, studying and controlling weak lensing systematic biases is becoming more critical \citep{rachelreview}.

One major source of observational systematics for WL is the point-spread function (PSF), which describes how a point source appears on the observed image. In the best case, the PSF is diffraction-limited, but in practice for ground-based surveys, it includes a dominant atmospheric contribution alongside optical aberrations and detector contributions.
While the atmospheric contributions to the PSF shape vary on short timescales, the CCD detectors have complex PSF shape systematics \citep{2018SPIE10709E..1LB} which are very similar every time a field on the sky is revisited, and thus do not average down over the course of the survey. Traditionally, the PSF is modelled empirically using images of stars (e.g., \citealt{psfex}) and then that model is used to infer the shapes of galaxies, which are tracers of the coherent weak lensing distortions (e.g., \citealt{great3}). It is necessary to use large numbers of stars in each CCD in order to adequately sample the PSF variations across the focal plane.  The shapes of brighter stars may be contaminated by flux-dependent detector effects; this is known as the ``brighter-fatter effect'' \citep[BFE;][]{antilogus}, which causes flux-dependent PSF model systematics, must be corrected in pixel processing and could be additionally mitigated via careful observing strategy choices that reduce the impact of the brighter-fatter effect in the final coadded image. 

PSF modelling is imperfect in practice, and errors in modelling the PSF lead to systematic biases on the cosmic shear signal, which is the two-point correlation function of the shear field produced by the large-scale structure (LSS) of the Universe;  see, for example, \cite{Paulin-Henrikkson+08} and \cite{rowe} for the formalism describing how PSF modelling errors propagate into the measured cosmic shear signals.  
A bias in the PSF model shapes would translate to a very significant additive bias in the shear power spectrum (e.g., \citealt{Paulin-Henrikkson+08} and \citealt{jarvis}).
While analyses of previous large-area surveys of galaxies must improve PSF modelling and interpolation algorithms to reduce the impact of PSF modelling errors on weak lensing measurements, the LSST survey provides an additional new option for systematics control: optimising the observing strategy. 

The main LSST survey is a Wide Fast Deep (WFD) survey \citep{Overview}, such that the majority of the LSST observing time will be spent carrying out a wide-area survey that is deep in limiting magnitude with many short observations. More specifically, the LSST is designed, according to the LSST Science Requirements Document (SRD) \citep{SRD}, 
to have a median of 825 visits across the 18~000 deg$^2$ footprint and across all of the $ugrizy$ bands. A visit is currently defined in the baseline strategy as two co-added 15-second exposures with a readout in between.
Like previous surveys, LSST will dither between observations at a given sky location, but unlike previous surveys, the LSST will have a unique combination of large-scale dithers and a large number of exposures at each point.  Thus, objects can be observed in significantly different positions in the focal plane due to offsets of telescope pointings (what we will call translational dithering), and with multiple angles due to offsets of the camera rotational angles (what we will call rotational dithering). These aspects of the observing strategy can be used in addition to traditional methods to mitigate weak lensing systematics. 
Related studies that explore possible translational dithers to determine how the LSST observing strategy can be used to reduce systematic errors in measurements of the large-scale structure 
have already been conducted (\citealt{carroll+14}, \citealt{humna}, \citealt{COSEP}).

In this paper, we study how different aspects of the observing strategy help mitigate the additive shear bias, and rank a set of simulated strategies based on their performance for WL systematics (while the statistical trade-offs are not addressed here). 
In Section~\ref{background}, we explain WL and its systematics in more detail; in Section~\ref{surveystrategy}, we present a representative set of LSST survey strategies and separately the three translational dithering strategies that are studied in this paper, although our method can be easily applied to new observing strategies in the future as they are released. In Section~\ref{method}, we present our methodology, using both a direct effect on the cosmic shear bias, and simpler statistical tests of uniformity; and in Section~\ref{results}, we analyze the results, and discuss them in the context of the 2018 call\footnote{\url{https://www.lsst.org/call-whitepaper-2018}; a full list of the white papers submitted in response to this call is available at \url{https://www.lsst.org/submitted-whitepaper-2018}} for proposals on optimising the LSST observing strategy for different science cases. 

\section{Background}
\label{background}
This section includes background information on weak lensing, the observable  quantities that are measured and used to constrain cosmological parameters, and the observational systematics that can be affected by different choices in observing strategy.

\subsection{Weak Lensing Measurements}

WL is a ubiquitous statistical effect that modifies the light profiles of galaxies as the light rays from those galaxies pass by other mass along the line of sight before they are observed. WL by the large-scale structure of the universe distorts the shapes and sizes of galaxies according to the distortion matrix:
\begin{equation}
\text{A} = \begin{pmatrix}
1-\kappa-\gamma_1 & -\gamma_2 \\
-\gamma_2 & 1-\kappa + \gamma_2
\end{pmatrix},
\end{equation}
where $\kappa$ is the convergence, a measure of the magnification due to WL, and $\gamma_1, \gamma_2$ are spin-2 shear, a measure of the shape distortion due to WL \citep{weaklensingreview}. The distortion matrix can be used to transform lensed coordinates into unlensed coordinates, such that for unlensed and lensed positions on the sky $\vec{x}_\text{u}$ and $\vec{x}_\text{l}$, $\vec{x}_\text{u} = A \vec{x}_\text{l}$ \citep{great3}.

The cosmic shear signal can be measured using shear-shear correlation functions between pairs of galaxies, as follows:
\begin{align}
\xi_+ (\theta) &= \mathbb{E} [\gamma \gamma^*]  (\theta) =  \mathbb{E}[ \gamma_t \gamma_t ] (\theta) + \mathbb{E} [\gamma_\times \gamma_\times ] (\theta),\\
\xi_- (\theta) &= R (\mathbb{E}[ \gamma \gamma ] (\theta) e^{-4i\phi}) =  \mathbb{E}[ \gamma_t \gamma_t ] (\theta) - \mathbb{E}[ \gamma_\times \gamma_\times ] (\theta),
\end{align}
where $t$ and $\times$ are the tangential and cross components of the shear, $\theta$ is the angular separation on the sky, $\phi$ is the polar angle \citep{weaklensingreview2}, $\mathbb{E} [ ] $ refers to the expected value and $R()$ refers to the real component.
We only consider $\xi_+$ in our analysis, because the biases on $\xi_-$ are much closer to 0, and so not much work is needed to mitigate them \citep{jarvis}.

It is clear from the definition of these correlation functions that any bias in the measured galaxy shears, either multiplicative or additive, will alter the measured shear correlation functions $\xi_{\pm}$. As we present more quantitatively in Section~\ref{method}, errors in the shape of the PSF model generate additive systematics, while errors in the size of the PSF model generate both multiplicative and additive biases in the cosmic shear correlation function.

While multiplicative biases in weak lensing shear are important and need to be carefully controlled, observing strategy cannot as easily mitigate coherent PSF size errors of a fixed sign (being a scalar, non-zero mean size errors will not average down with translationally or rotationally dithered observations), so in this paper we focus on capturing the impact of observing strategy on PSF model shape errors (and therefore on additive systematics in WL).

\subsection{Weak Lensing Systematics}
\begin{figure}
    \centering
    \includegraphics[width=\columnwidth]{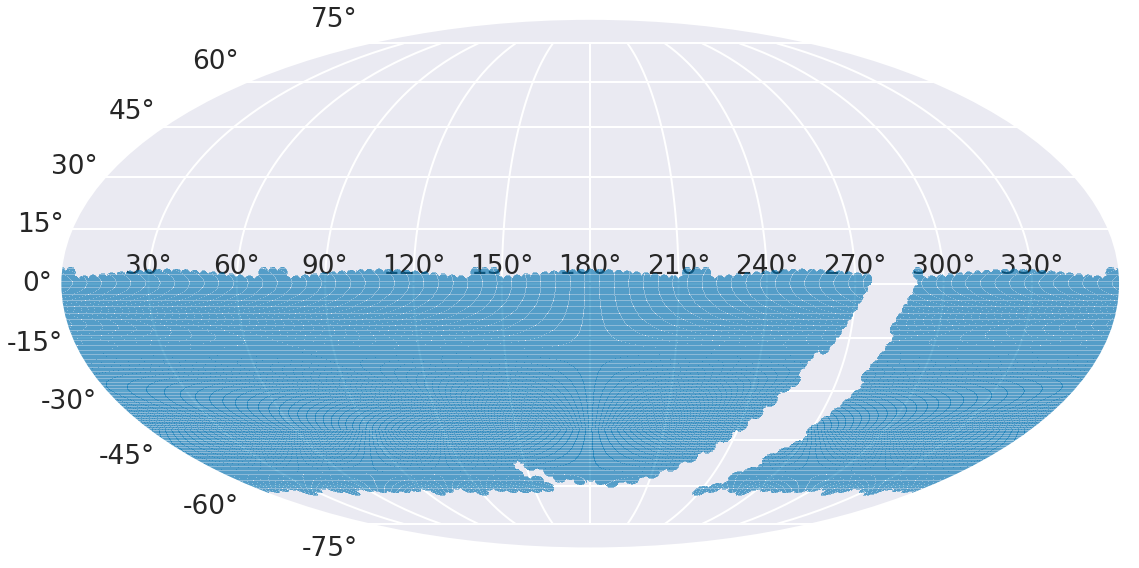}
    \caption{The right ascensions and declinations (survey footprint) of 444,867 focal plane field positions (without translational dithering) for $i$-band observations in the entire main WFD survey during the 10-year survey in \texttt{baseline2018a}. 
    \label{fig:baselinefieldpos} }
\end{figure}

\begin{figure}
    \centering
    \includegraphics[width=0.35\textwidth]{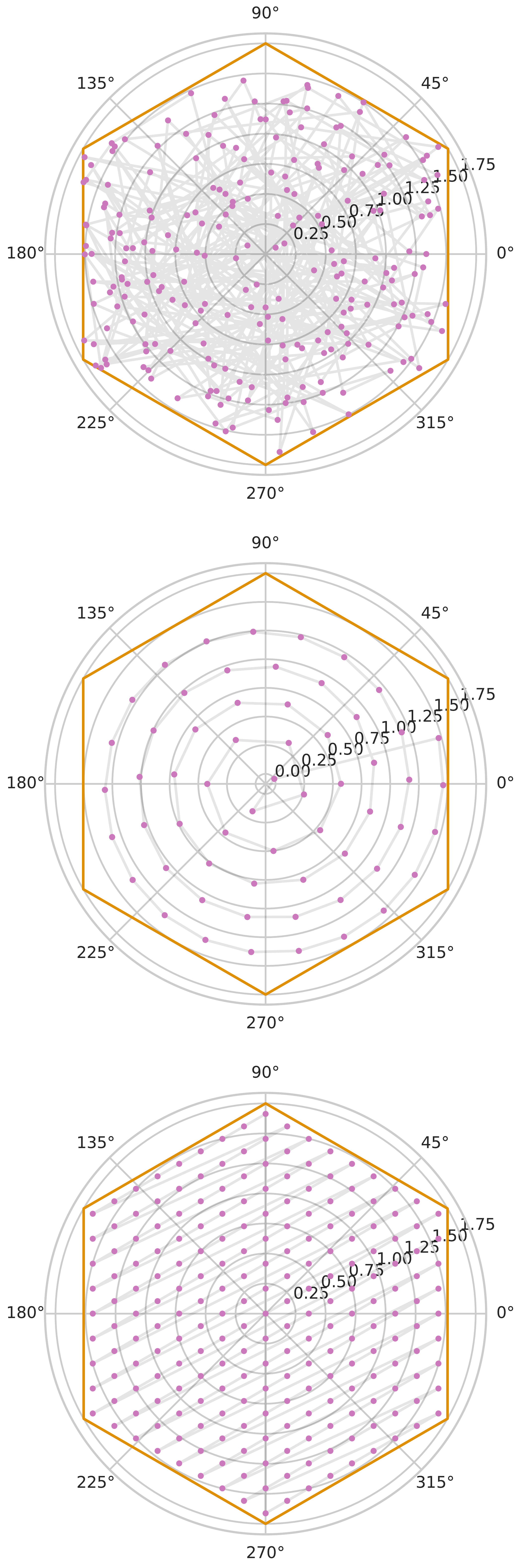}
    \caption{
    The translational dither patterns that are studied in this paper. These are, from top to bottom: random, spiral, and hexagonal. The pink points show the dithered positions, stepped through sequentially according to the gray lines. The plots are scaled to the size of the FOV of the LSST. These plots are generated using the Metrics Analysis Framework  \protect\citep[MAF;][]{MAF} following the approach defined in  \protect\cite{humna}. Unlike the random strategy, the hexagonal and spiral strategies are sequential and start to repeat at large numbers of dithers, resulting in an apparently lower unique pointing density even for the same number of dithers planned.} 
    \label{fig:ditherpatterns}
\end{figure}

\begin{figure}
    \centering
    \includegraphics[width=0.45\textwidth]{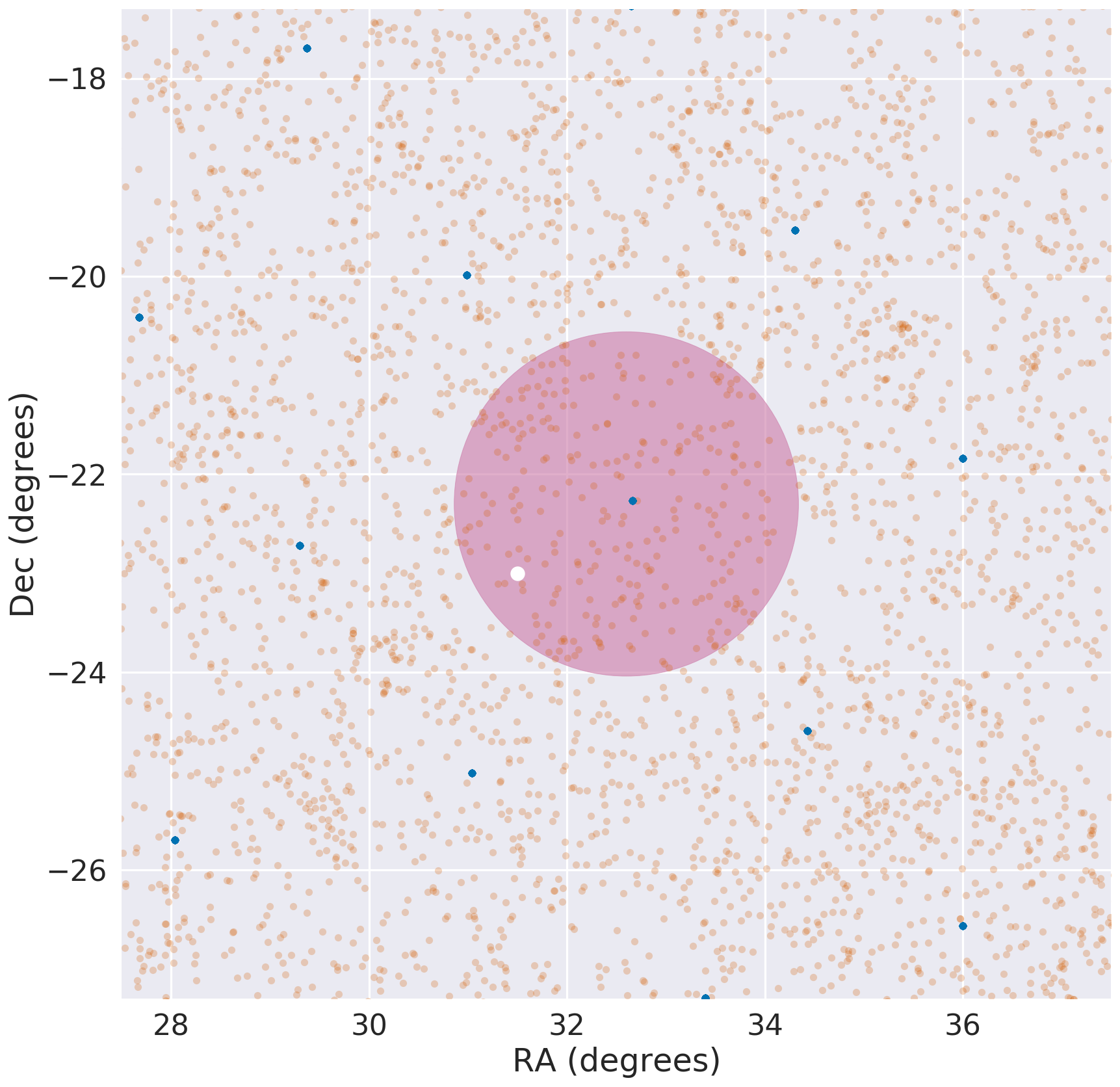}
    \caption{Illustration of the effect of dithering: The red circle represents the field of view of the LSST camera for a single exposure, with its centre indicated by one of the blue dots (indicating field positions). An object, indicated by the white circle, would only be imaged at a single position in the focal plane if no translational dithering is carried out.
    With random translational dithering around the fixed field positions (blue dots), the object would be imaged at $\sim$200 different positions within the focal plane, corresponding to the total number of visits in the $i$-band, 
    with centres of those exposures indicated by the orange dots. The plot was made using the field positions in the reference observing strategy (\texttt{baseline2018a}), with random dithering applied. \label{fig:numberofdithers} }.
\end{figure}

Several sources of systematics can cause multiplicative or additive biases in the cosmic shear signal. These sources can be either theoretical, astrophysical, or observational. Theoretical sources of systematics include failure of the Limber approximation, likelihood function inaccuracies, and covariance misestimation (e.g., \citealt{theoreticalsystematics1}, \citealt{theoreticalsystematics2}, \citealt{theoreticalsystematics3}). Astrophysical systematics include, for example, intrinsic alignments (e.g., \citealt{IA1}, \citealt{IA2}, \citealt{IA7}, \citealt{IA6}), the fact that galaxies are not oriented randomly throughout the universe even in the absence of lensing. Precise theoretical models of intrinsic alignments are required to turn WL measurements into cosmological parameters. Observing strategy cannot impact theoretical systematics, and the impact of observing strategy on astrophysical systematics is accounted for using statistical forecasting. We, therefore, focus on the impact of observing strategy on observational systematics. The remainder of this subsection is dedicated to providing background on  observational systematics, and their effect on the cosmic shear signal.

When an observatory measures an image of a point source, it observes an extended and potentially complex light profile, due to the effect of the atmosphere, the optics (and optical aberrations), the CCD sensors, and the electronics. Atmospheric effects are mainly due to turbulence and vary stochastically with time and spatially across the focal plane (e.g., \citealt{atmosphericpsf}). The effects of telescope optics include not only the obscured Airy diffraction pattern, but also  aberrations that can be expressed in the form of Zernike polynomials such as coma and astigmatism (e.g.,  \citealt{opticalpsf}, \citealt{roodman+2014}). Finally, detector effects 
(such as the ones due to charge transport asymmetry) are not stochastic; they remain the same when a field is revisited, and contribute to the PSF shape integrated over all visits unless removed by correction in pixel processing and camera rotation during observing.  
Most of these effects, when combined, can be described using a PSF. The galaxy images are convolved with the PSF, which must be modelled carefully to remove the effect of the PSF and  recover unbiased weak lensing shear estimates.

Most PSF modelling techniques use images of stars as an effective PSF at their observed positions, and interpolate to get a PSF model that can be evaluated at any point within the focal plane.  Other techniques model optical aberrations physically and are usually more appropriate for optics-limited space telescopes. The LSST PSF modelling strategy may incorporate elements of both methods \citep{rachelreview}. 
While methods to correct for the impact of the PSF on the galaxy shapes  expect PSF modelling to be carried out perfectly, in practice no PSF modelling method is perfect \citep{kitching+2013}. Reasons for PSF model insufficiency include the low density of stars, interpolation techniques that do not fully describe the physical processes governing the variation of the PSF across the focal plane, and detector effects such as the brighter-fatter effect \citep{antilogus}, wherein the PSF measured from bright stars appears larger than it actually is.  The physical origin of this effect is that the electrons in a pixel spill out to neighboring pixels due to electrostatic repulsion, violating the linear relationship between electron counts and exposure time expected for CCDs.  Source galaxies used in WL studies are typically faint, so a PSF model inferred from bright stars without correction for the brighter-fatter effect misrepresents the actual PSF. Current methods to correct for the brighter-fatter effect, such as the one demonstrated using the Hyper Suprime-Cam (HSC; \citealt{hsc}) survey data in \citealt{hscShearCatalog} are not exact, so some residual brighter-fatter effect may still be expected to contaminate PSF models.  Additionally, when the accumulated charge is transferred across the CCD pixels during readout, some amount of charge is lost in the process; this effect is known as charge transfer inefficiency \citep[CTI;][]{cti}. This effect introduces a residual signal along the charge transfer direction, altering the perceived shapes of stars and galaxies. While BFE is a larger effect than CTI for most LSSTCam CCDs\footnote{Aaron Roodman, private communication, Aug 9th, 2020.}, their relative magnitude on Rubin Observatory CCDs after applying software corrections is still being studied.

PSF modelling imperfections can often imprint a coherent PSF shape bias in a specific direction in the plane of the camera. Simulations of LSST observing using laboratory measurements on LSST CCDs reveal multiple PSF shape systematics which can only partially be removed in pixel processing \citep{2018SPIE10709E..1LB}.
Observing strategy can, therefore, be an effective way to average down this bias significantly in addition to what can be gained with improved software for PSF modelling or for removing detector effects in the initial pixel processing steps. Examples of systematics with coherent special directions include radially-oriented residuals within the focal plane due to PSF modelling errors, as has been observed in e.g., \cite{jarvis} and \cite{hscpipeline}. Residuals associated with the orientation of the camera focal plane due to CCD fixed-frame distortions and differential chromatic refraction effects are discussed in \cite{COSEP}: these systematics were found to be optimally suppressed for observing strategies with uniform distributions (over the range $[0, \pi]$ radians) of parallactic angle and the angle between the +y camera direction and the North (referred to as rotSkyPos).  We extend this analysis to the direction of CCD charge transfer, which would be horizontal or vertical, to account for physical effects such as the brighter-fatter effect and charge transfer inefficiency, which could result in an additive shear systematic error. 

\section{LSST Observing Strategy}
\label{surveystrategy}
In this section, we describe the tools used for simulating and analyzing LSST observing strategies, and describe the survey simulations that are used for our analysis. 
\subsection{LSST Operations Simulator (OpSim)}
The LSST cadence and observing strategy have not yet been decided \citep{COSEP}\footnote{\url{https://github.com/LSSTScienceCollaborations/ObservingStrategy}}. 
The LSST Operations Simulator (OpSim\footnote{\url{https://www.lsst.org/category/operations-simulation}}; \citealt{Opsim}) can be used to simulate the effects of survey strategies on survey parameters. OpSim combines science program requirements, telescope design mechanics, and modelling of environmental conditions to provide a framework for operational simulations which return the parameters of the survey that do not specifically require image simulations, such as exposure positions, airmass values, the position of the moon at each exposure and filter change.

\subsection{Metrics Analysis Framework (MAF)}
\label{sec:maf}
The Metrics Analysis Framework (MAF\footnote{\url{https://www.lsst.org/scientists/simulations/maf}}; \citealt{MAF}) is a tool to assess the impact of observing strategy on particular science cases. MAF is an object-oriented analysis framework that facilitates implementation of metrics that use the output of operational simulations (e.g., OpSim runs) to consistently generate metrics. Two weak lensing-related metrics are already included in MAF, as described in Section 9.3 of ~\cite{COSEP}: those are the AngularSpread\footnote{\url{https://github.com/LSST-nonproject/sims_maf_contrib/blob/master/mafContrib/angularSpread.py}} metric and KuiperMetric.  These metrics measure the uniformity of the distribution of the rotational angle of the camera. They quantify how well a certain observing strategy averages down the additive shear systematics induced by non-uniformity of parallactic angle and rotational sky position.  

A metric from the present work is incorporated into MAF\footnote{\url{https://github.com/lsst/sims_maf/blob/master/python/lsst/sims/maf/metrics/weakLensingSystematicsMetric.py}}; this metric is discussed in Section~\ref{sec:metric}. This metric should be sufficient, under some assumptions, to represent the information from the full analysis of the additive bias on the cosmic shear for a given dithering strategy, and can be used to compare different observing strategy choices once a dithering strategy is set.

\subsection{Survey Strategies Studied}

\label{sec:strats}
\begin{table*}
    \caption{Summary of the observing strategies (OpSim runs) that are used in this paper. 
    }
    \begin{tabular}{ll}

        \label{tab:runs}
        Strategy &  Description \\ \hline \hline
        \texttt{baseline2018a} & LSST official baseline strategy  \\ 
        \texttt{pontus\_2002} & 24,700 deg$^2$ footprint (instead of 18,000 as in \texttt{baseline2018a}) \\
        \texttt{kraken\_2042} & 1$\times$30s visits (instead of 2$\times$15s as in \texttt{baseline2018a)} \\
        \texttt{pontus\_2489} & 1$\times$20s visits in $grizy$ and 1$\times$40s in $u$ (instead of 2$\times$15s visits as in \texttt{baseline2018a}) \\
        \texttt{colossus\_2664} &  considers the Galactic plane part of the WFD survey (spends more time on it than \texttt{baseline2018a})
    \end{tabular}
\end{table*}

An official OpSim reference simulated survey, \texttt{baseline2018a}\footnote{\url{http://astro-lsst-01.astro.washington.edu:8080}, \url{http://ls.st/doc-28382}} assigns field positions covering the 18,000 deg$^2$ of sky in declinations ranging between $-62^{\circ}$ and $+2^{\circ}$, with at least 825 visits per field across all of the six $ugrizy$ filters. 
In addition to the main WFD survey, \texttt{baseline2018a} schedules other science proposals (mini-surveys): the Galactic plane, 5 deep drilling fields (DDFs), the north ecliptic spur, and the south celestial pole. 
Fig.~\ref{fig:baselinefieldpos} shows the focal plane centres in \texttt{baseline2018a} without taking into account dithering.  

Four other OpSim strategies that were made available as a part of the call for observing strategy white papers by the LSST Project in 2018 to the LSST science community to help define the observing strategy, are also studied in this paper.  We selected these strategies among those provided by the Project as representative examples of simulations that exhibit features of particular relevance to weak lensing systematics mitigation. 
These features are: a large area (\texttt{pontus\_2002}), shorter visits (\texttt{pontus\_2049}), single-exposure visits (\texttt{kraken\_2042}). The baseline strategy defines a visit as two 15-second exposures with a readout in-between, which we refer to as 2$\times$15s; we will refer to a single-exposure 30-second visit as 1$\times$30s. 
A noteworthy category of strategies is rolling cadence strategies. These focus on a limited band of declination for a period of time, and then move on to other declination bands, leading to a shorter interval between repeated visits for the declination band that is being observed at any given time.  These strategies are particularly of interest for transient science, due to the better sampling of light curves. Due to the lack of rolling strategy OpSim runs that are comparable with the ones selected for investigation in this paper, we are not including rolling cadence strategies in this work. 
Table~\ref{tab:runs} summarises the strategies considered in this work.

Observing strategy simulations have also been generated using tools other than OpSim, and have been studied and ranked along with other strategies in \citet{OSTFWhitepaper}; in particular, these are the feature-based strategy  \texttt{slair} \citep{paper:slair}, and \texttt{ALT\_Sched} \citep{paper:ALTSched}. However, we excluded these strategies from this work because (a) they did not provide new information that influences our results, and (b) they use dramatically different algorithms than OpSim, which complicates interpretation of the results. Some beneficial aspects of the algorithmic changes of these runs have been incorporated in later OpSim development. 

For the majority of the analysis, we made cuts on the dust-corrected minimum co-added depth. The cut at the end of the 10th year of the survey (Y10) excludes regions with depth shallower than point-source magnitude $i=26$ mag, corresponding to `gold sample' galaxies with extended-source magnitude $i<25.3$ mag, based on an approximate conversion between the magnitudes, following the DESC Science Requirements Document \citep{DESCSRD}.  At the end of the first year of the survey (Y1), the cut excludes regions shallower than point-source magnitude $i=24.5$ mag, assuming the survey limiting magnitude shifts by 2.5 times the base-10 logarithm of the observing time.  A more detailed depth optimization would be valuable for future work. 
In addition, we made cuts based on extinction, considering only areas with values of reddening $E(B-V) < 0.2$. This cut eliminates areas of high extinction and high dust uncertainties \citep{ebv}; practically, this cuts out the majority of the Galactic equator. More quantitatively, applied on \texttt{baseline2018a}, it reduces the area of the 10-year survey from 18,040 to 14,695  deg$^2$; after that, a co-added depth cut of $i > 26$ mag only slightly reduces it further to 14,691 square degrees. 

\begin{figure}
    \centering
    \includegraphics[width=0.45\textwidth]{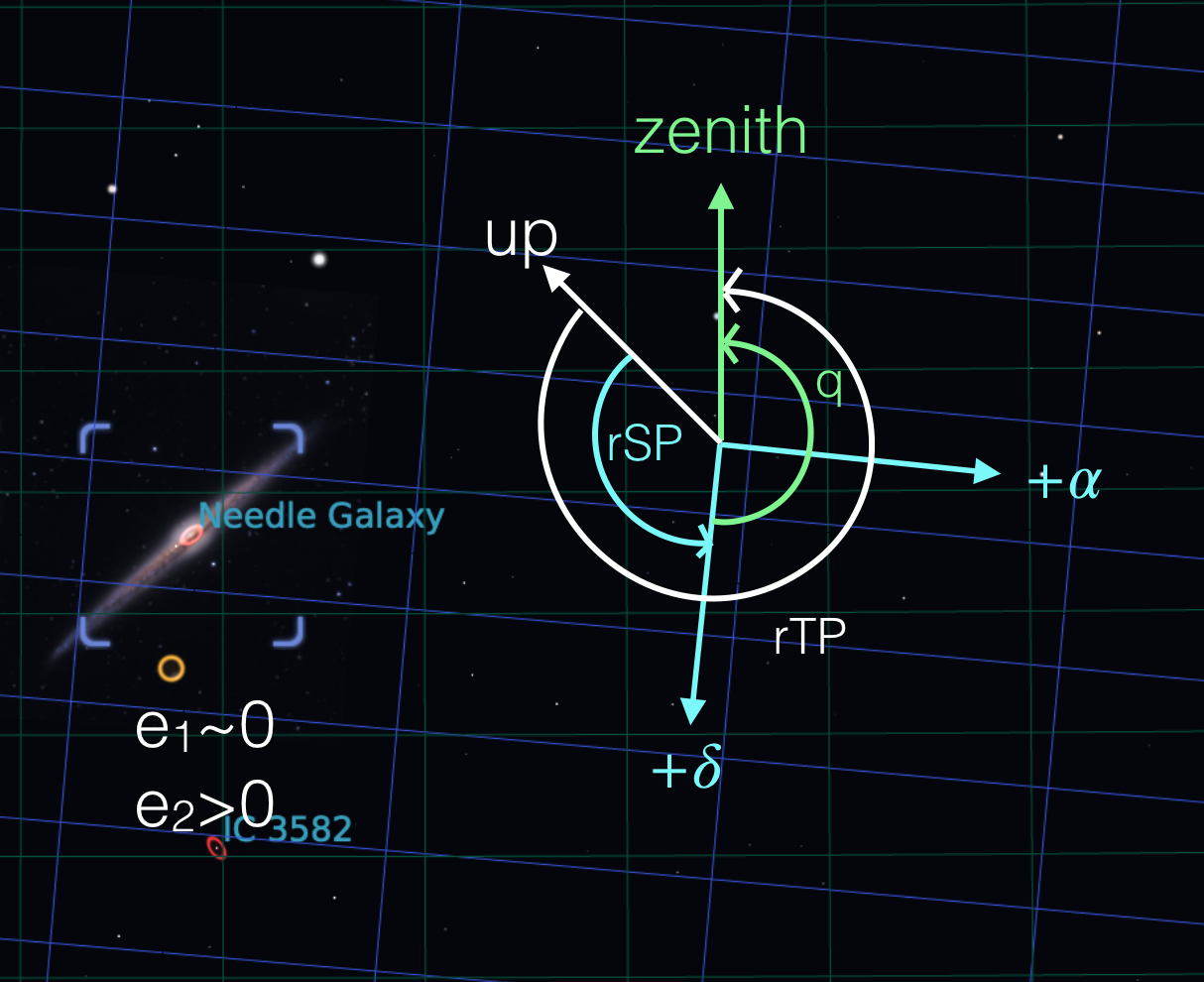}
    \caption{A summary of the angles mentioned in this paper. rSP and rTP refer to rotSkyPos and rotTelPos respectively. The figure indicates the right ascension and declination axes defined in the equatorial coordinate system indicated by the blue grid. The green grid corresponds to the horizontal coordinate system (with the Zenith being aligned with the altitude). Up is the +y direction in a single exposure, and therefore, what we refer to as axis-parallel models are perpendicular (or parallel) to Up, while radial models are radial in the plane of the figure. Radial is defined as purely pointing towards the centre and changing with radius. Finally, the parallactic angle $q$ is the angle between rTP and rSP. The figure also illustrates $e_1 = |e| \text{cos}(2\vartheta)$ and $e_2 = |e| \text{sin}(2\vartheta)$, where $|e| = (a^2-b^2)/(a^2+b^2)$.  Here $a$ is the semimajor axis,  $b$ is the semiminor axis, and $\vartheta$ is the angle between the semimajor axis and the $+\alpha$ direction in the equatorial coordinate system.}
    \label{fig:angles}
\end{figure}

\subsection{Dithering}
We use two types of dithering in this paper. First, we incorporate translational dithering per visit into the strategies, and apply it to each field position, using one of three different translational dithering patterns: hexagonal, random, and spiral, all shown in Fig.~\ref{fig:ditherpatterns}. It is possible to use different dithering timescales (e.g., per-night dithering), and the impact on changing the timescale on our results will be discussed in this paper. The result of random dithering, as an example, is illustrated in Fig.~\ref{fig:numberofdithers}. Random dithering in particular refers to choosing a number of offsets at random (constrained within the size of the FOV), and applying them to the undithered field position. Dithering algorithms are used as implemented in MAF. 

Secondly, we apply rotational dithering at random between $[-90, 90]$ degrees at every filter change, to satisfy the physical restriction that the camera can only rotate within that range due to the camera's cable wraps. Filter changes also require a reset on the camera angle, and take four times as long as a rotational dither, so using filter changes as an opportunity to dither rotationally reduces overheads compared to an approach that considers separate filter changes and rotational dithers \footnote{\url{https://github.com/lsst-pst/survey\_strategy/blob/master/Constraints.md}}. The camera naturally rotates on small scales when tracking the sky, but we ignore this slewing. The reason behind this is that a more accurate rotational dithering algorithm needs to be used during the operation of the telescope (or when running the simulation) rather than afterwards in post-processing, so the limits on camera rotation angle are respected. Therefore, ignoring this additional slewing is necessarily more conservative than otherwise. A more sophisticated rotational dithering algorithm (e.g., one that aims to homogenise the imaging quality throughout the sky, taking into account seeing, airmass, etc.) can be implemented in the future, but this cannot be retroactively implemented within OpSim runs, and so it is beyond the scope of this paper. 

\section{Method}
\label{method}

\begin{figure}
\centering
\includegraphics[width=\columnwidth]{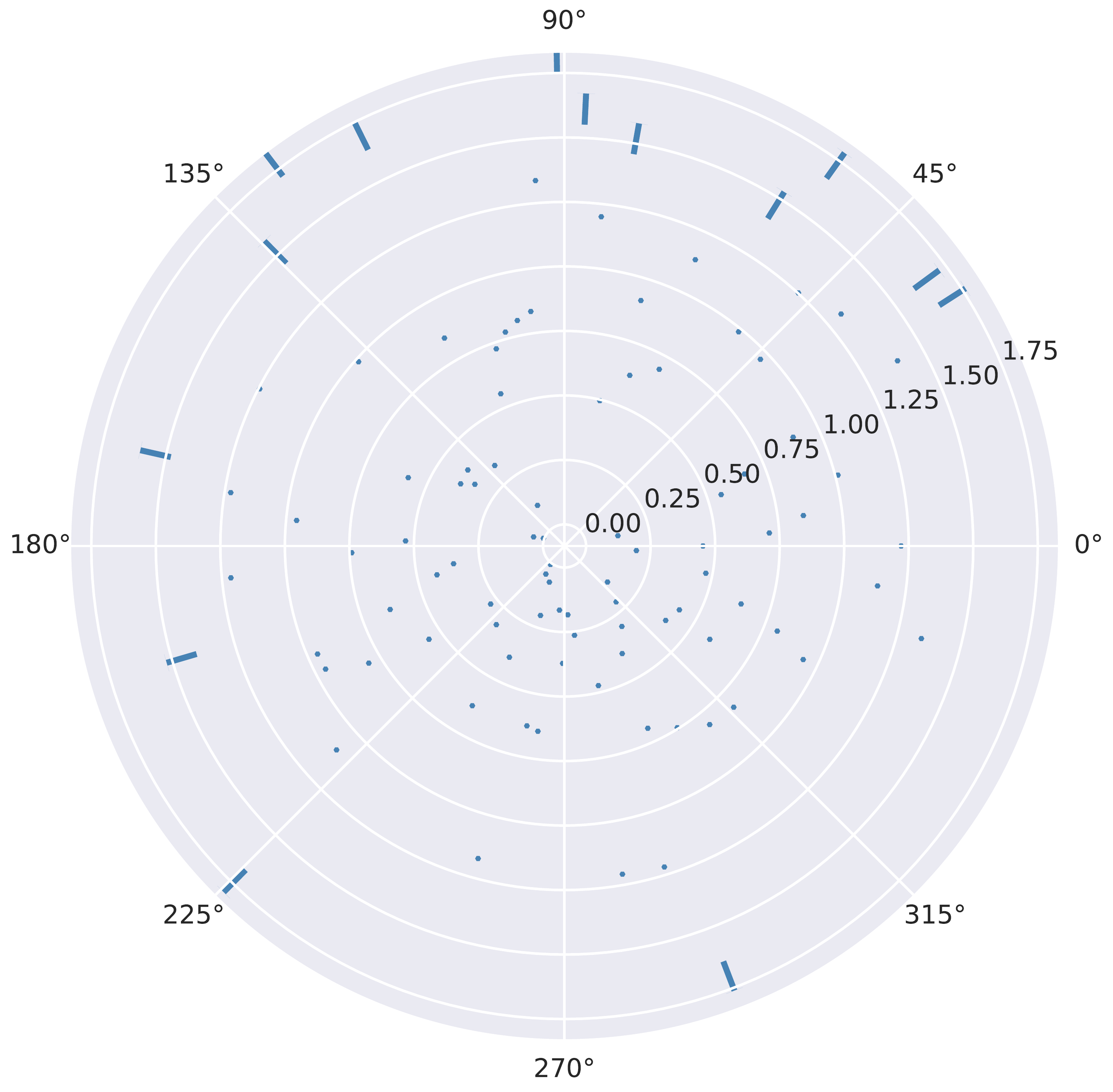}

\includegraphics[width=\columnwidth]{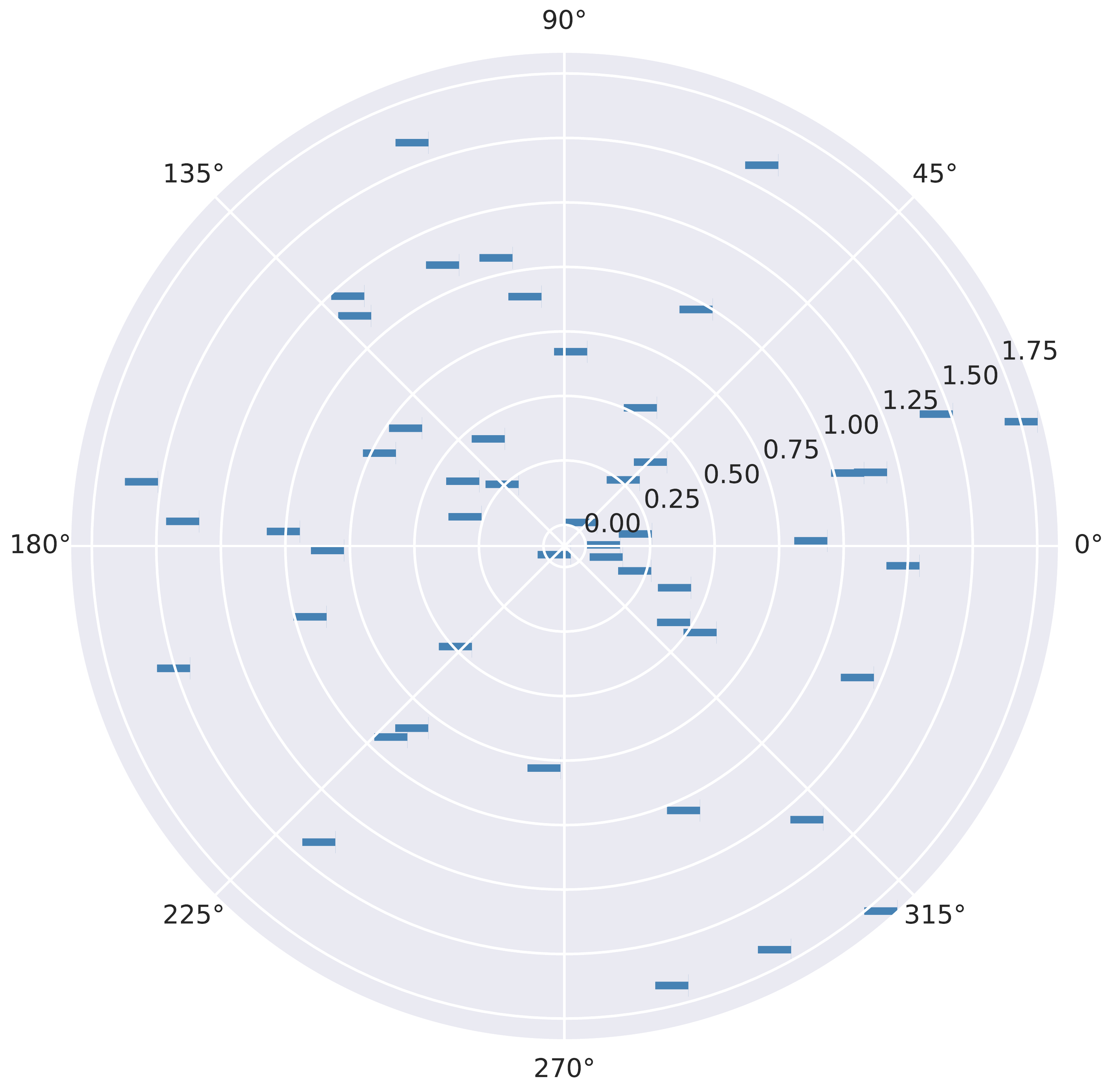}

\caption{\label{radial-and-horizontal} The radial (top) and horizontal (bottom) toy models for the PSF model shape errors due to PSF modelling imperfections and residual CCD charge transfer biases as explained in Sections~\ref{sec:ccd} and~\ref{psfmodelingerrors}. The orientations of the line segments represent the shear angle at their positions with respect to the centre of the focal plane, and their lengths represent the magnitude of the shear. The magnitudes of the shears are chosen such that the average residual ellipticity (apart from inner 80\% excluded areas -- see text for details) matches the values described in Sections~\ref{sec:ccd} and \ref{psfmodelingerrors}.}
\end{figure}

To define metrics for weak lensing additive systematic biases, we start by modelling the positions of a large number (here we use 100,000) stars randomly sampled within the usable area. We define the usable area as being the WFD area for the observing strategy in question, after placing a cut based on the co-added depth and extinction, as described in Section~\ref{sec:strats}. 
The distributions of properties of individual visits in the $i$-band and co-added properties for these stars are then used for a variety of systematics tests described in the subsections below. We define toy models for the true and estimated PSF size and shape within a single exposure as a function of position within the focal plane as described in Subsections~\ref{sec:ccd} and~\ref{psfmodelingerrors}. We present a formalism for combining them in  Subsection~\ref{sec:momentspace}. We then propagate the effects of PSF systematics onto two-point correlation functions used to constrain cosmological parameters in Subsection~\ref{cosmoprop}.  

Since observational weak lensing additive shear systematics are associated with special directions (typically, though not always, in single exposures), we can test whether specific special directions are being imaged uniformly when considering all exposures. The following subsections detail specific spatial patterns associated with particular sources of systematic uncertainty, and the uniformity tests that we use to rank three different dithering strategies and five observing strategy simulations with regard to how well they average down certain systematics. 

Fig.~\ref{fig:angles} summarises the angles referred to in the subsections below.

\subsection{Axis-parallel model}
\label{sec:ccd}

Several detector non-idealities induce systematics in the shear signal. The brighter-fatter effect can be correlated with the charge-transfer direction, leaving PSF model shape errors due to residual brighter-fatter effect along this direction (\citealt{rachelreview}); for example, this can be a purely horizontal or vertical residual. In addition, CTI also leave a residual along the charge transfer direction, i.e. leaving a horizontal or vertical residual in the image plane (see e.g., \citealt{ctifig}). Translational dithering will not average down this systematic bias (the residuals will look the same after any translational dithering). We therefore test the impact of a combination of translational and rotational dithers. 

\cite{2018SPIE10709E..1LB} suggest that more realistic LSST-specific residual due to CCD effects will be a combination of vertical and horizontal (due to the combination of BFE, CTI, and CCD output amplifier response); therefore, our horizontal-only model is necessarily more conservative than this combination of effects.

For a statistical uniformity test, we use a Kuiper test \citep{kspaper} to compare the distribution of charge transfer angles with respect to the centre of the focal plane for each star against a uniform distribution. In particular, for each dithered position, the Kuiper test returns a D-statistic, defined as the distance between (a) the empirical distribution function of angles between +x and the lines connecting a set of observed stars to the centre of the focal plane and (b) a specific reference cumulative distribution function (in our case the uniform cumulative distribution function). To more quantitatively forecast the effect on cosmological observables, we use a toy model of completely horizontal PSF model residuals (before rotational dithering), with $ |e_\text{PSF}|= \sqrt{ e_1^2 + e_2^2} = 0.05 $, and the difference between the true and estimated PSF model, $\Delta e = \hat{e}_\text{PSF} - e_\text{PSF} = 0.0015$. The horizontal model is illustrated in the bottom panel of Fig.~\ref{radial-and-horizontal}. At each dither position or centre of the focal plane for each exposure, we find all simulated stars visible in that exposure and apply this toy model for the PSF shape and PSF model shape residual in that exposure. 

It should be noted that what we refer to as rotSkyPos in Fig. \ref{fig:angles} is the angle that is relevant to the systematics that we are concerned about (and the one that we should isotropise to average down the axis-parallel systematics), since this is the angle that is between Up in the focal plane (a reference angle defined in the focal plane coordinate system) and North (the reference angle for the orientation of galaxies on the sky). rotSkyPos and rotTelPos (the latter defined as the angle between up in the focal plane and zenith) are related via the parallactic angle, which has been studied in the context of other systematics (e.g., differential chromatic refraction) in \cite{COSEP}.

\subsection{Radial model}
\label{psfmodelingerrors}

It has been empirically observed in previous surveys (e.g., HSC) that errors arising from imperfect PSF modelling using the current state-of-the-art PSF modelling algorithms, such as PSFEx \citep{psfex}, exhibit radial patterns that significantly increase near the edges of the focal plane (as can be seen in, e.g., Fig. 9 in \citealt{jarvis} and Fig. 9 in \citealt{hsc}). These may arise due to difficulty in modelling the complex and strongly-varying optical PSF component in those regions.
While newer PSF modelling algorithms (e.g., Piff\footnote{\url{https://github.com/rmjarvis/Piff}}) are under development and may improve upon the current state of the art, they are not yet sufficiently well demonstrated, and radial PSF residuals are likely to persist with any method at some level since real PSFs typically have a strong radial component at the edges of the focal plane. Hence, it is worth investigating the impact of observing strategy on  radial PSF model shape errors near the edge of the focal plane.  
The special direction associated with additive systematics due to this type of PSF modelling error, therefore, points towards the centre of the focal plane. Thus, assuring the uniformity of the distribution of angles between the line connecting the observed objects to their respective centres of focal planes in individual exposures and +x (perpendicular to the `up' direction in Figure~\ref{fig:angles}) is needed to reduce systematic uncertainties through observing strategy. 
For a statistical uniformity test, we use the Kuiper test to assess the uniformity of the distribution after dithering. To measure the effect on cosmic shear, we use a toy model that assumes a perfect PSF model for stars that are within 80\% of the radius of the FOV; outside of that range, the radial PSF model shape and its error are set to $\mathbb{E} [\Delta e_{\text{radial}} ] = 0.005$ with $\mathbb{E} [ e_{\text{radial}} ] = 0.08$, where $\Delta e = \hat{e}_\text{PSF, radial} - e_\text{PSF, radial} $ is the difference between the true and estimated PSF model. This toy model is illustrated in the top panel of Fig.~\ref{radial-and-horizontal}.

\subsection{Averaging Across Exposures}
\label{sec:momentspace}
Due to the combination of rotational and translational dithering, every star will be imaged from many positions within the focal plane with different orientations for the PSF model shape residuals, which will allow us to effectively average down the residual ellipticities. Given that the shape and size of a star or a galaxy are defined by second moments of the light profile, we first convert our toy PSF model shapes and residual ellipticities to second moments.  
The residuals are defined as:
\begin{equation}
\begin{aligned}
\delta e_1 =  \hat{e}_1^{\text{PSF}} - e_1^{\text{PSF}},
\delta e_2 =  \hat{e}_2^{\text{PSF}} - e_2^{\text{PSF}}.
\end{aligned}
\end{equation}

The shape and size of an object are, by definition: 
\begin{equation} \label{moments1} e_1 = \frac{M_{xx} - M_{yy}}{\mathrm{Tr}M}, \end{equation}
\begin{equation} \label{moments2} e_2 = \frac{2M_{xy} }{\mathrm{Tr}M},
\end{equation}
where $\mathrm{Tr}M = M_{xx} + M_{yy}$ is the trace of $M$. 
Using Equations \eqref{moments1} and \eqref{moments2}, we get:
\begin{equation}
M = \frac{\mathrm{Tr}M}{2} \begin{bmatrix}
     1 + e_1       &  e_2  \\
     e_2       &  1 - e_1 \\
     
\end{bmatrix}.
\end{equation}

In a coadded image constructed based on the weighted mean of image intensities in individual exposures, the intensity and hence the second moments add linearly. The weight function in the coaddition typically relates to factors such as the sky background noise, and does not generally correlate with the PSF shape. This allows us to assume that all epochs get the same weight in our toy model.
Hence, assuming no astrometric errors, we take the arithmetic mean of each matrix element $M_{ij}$ to get $\mathbb{E} [M] = N^{-1} \sum_{l=1}^N M_{ij,l}$ for all exposures $l$ at a certain sky location, and then we can go back to ellipticity space using equations~\eqref{moments1} and~\eqref{moments2}.

The size parameter Tr$M$ can be an arbitrary number, as it does not affect the values of the ellipticities throughout the process described here.

\subsection{Effect on Cosmological Measurements}
\label{cosmoprop}

We use the $\rho$-statistics, as in \cite{rowe} and \cite{jarvis}, to propagate the ellipticity and size residuals due to the PSF modelling errors. The $\rho$ statistics are defined as 

\begin{equation}
\begin{array}{r@{}l}
\rho_1 (\theta) &{}= \mathbb{E}[ \delta e_{\text{PSF}}^*(x) \; \delta e_{\text{PSF}} (x+\theta) ], \\
\rho_2 (\theta) &{}= \mathbb{E}[ e_{\text{PSF}}^* (x) \; \delta e_{\text{PSF}} (x+\theta) ], \\
\rho_3 (\theta) &{}= \mathbb{E}[ \left( e_{\text{PSF}}^*   \dfrac{\delta T_{\text{PSF}}}{T_{\text{PSF}}}\right)(x) \; \left(e_{\text{PSF}}   \dfrac{\delta T_{\text{PSF}}}{T_{\text{PSF}}}\right)(x+\theta)], \\
\rho_4 (\theta) &{}= \mathbb{E}[ \delta e_{\text{PSF}}^* (x)\;\left( e_{\text{PSF}} \dfrac{\delta T_{\text{PSF}}}{T_{\text{PSF}}}\right)(x+\theta)], \\
\rho_5 (\theta) &{}= \mathbb{E}[ e_{\text{PSF}}^* (x)\; \left(e_{\text{PSF}} \dfrac{\delta T_{\text{PSF}}}{T_{\text{PSF}}}\right)(x+\theta) ], \\
\end{array}
\end{equation}
where $T_\text{PSF} = \mathrm{Tr}M $ is the PSF model trace, and $\delta T_\text{PSF}$ is the error in the PSF model trace.

These $\rho$-statistics are correlation functions of different combinations of PSF model shapes, shape residuals, and size residuals, defined because they can be directly related to the total additive bias on the cosmic shear. Given the $\rho$ statistics, the bias in the cosmic shear signal is of the order:

\begin{equation}
\begin{aligned}
\delta \xi_+ (\theta) ={} & 2 \mathbb{E}\left[ \frac{T_{\text{PSF}}}{T_{\text{gal}}} \frac{\delta T_{\text{PSF}}}{T_{\text{PSF}}} \right] \xi_+(\theta) \\
&{} + \mathbb{E}\left[ \frac{T_{\text{PSF}}}{T_{\text{gal}}} \right]^2 \rho_1(\theta) 
 - 2 \alpha\mathbb{E}\left[\frac{T_{\text{PSF}}}{T_{\text{gal}}}\right] \rho_2(\theta) \\
&{} + \mathbb{E}\left[ \frac{T_{\text{PSF}}}{T_{\text{gal}}} \right]^2 \rho_3(\theta) 
 + \mathbb{E}\left[ \frac{T_{\text{PSF}}}{T_{\text{gal}}}\right]^2 \rho_4(\theta) \\
&{} - 2 \alpha\mathbb{E}\left[\frac{T_{\text{PSF}}}{T_{\text{gal}}}\right] \rho_5(\theta), \\
\end{aligned}
\end{equation}
where $T_\text{gal}$ is the true galaxy trace, and  $\alpha$ measures the leakage of the PSF shape into the galaxy shapes, which we take as $0.01$, consistent with the current state-of-the-art methods \citep{troxel2018}. We assume there are no PSF model size errors, which allows us to drop the terms containing $\rho_3, \rho_4$ and $\rho_5$, after checking empirically that for a typical value of $\frac{\delta T_{PSF}}{T_{PSF}} = 0.001$, and under the assumption that they do not correlate strongly with PSF shape or PSF model shape errors, including those three terms changes the value of $\delta \xi_+$ by less than 2\%. 

We use a sample of galaxies from the COSMOS catalog\footnote{\url{https://github.com/GalSim-developers/GalSim/wiki/RealGalaxy-Data}} \citep{great3} with limiting $i$-band magnitude of $25.2$, which is input to  GalSim\footnote{\url{https://github.com/GalSim-developers/GalSim}} \citep{galsim} to calculate the ratio $\mathbb{E} \left[ T_{\rm PSF}/T_{\rm gal} \right] $ as follows: first, for every galaxy in COSMOS, we use GalSim to simulate a parametric model of it using Sersic profiles as described in \cite{great3} of it. 
We then calculate its adaptive moments (weighted second moments for which the weight function is an elliptical Gaussian that is iteratively adjusted to match the moments of the objects being measured). We also draw a FWHM value for the PSF from a log-normal distribution with a median of 0.6 arcsec and standard deviation of 0.1 arcsec.  These are the best-fit parameters of  the distribution of PSF FWHM values measured at the Cerro Pach\'{o}n site using a Differential Image Motion Monitor and corrected using an outer scale parameter of $30$~m.\footnote{\url{https://www.lsst.org/scientists/publications/science-requirements-document}} To make the model more realistic, we shift it from a wavelength of $500$~nm (that is provided in the LSST SRD) to $800$~nm (corresponding to the $i$-band that we are working in), assuming a power-law wavelength dependence for the PSF FWHM, with an index of $-0.2$. We then add 10\% to the PSF FWHM to account for non-atmospheric PSF effects (10\% of the atmospheric contribution is the upper limit for the non-atmospheric contribution to the PSF size as specified in the LSST SRD). We use GalSim to draw a Kolmogorov profile with this FWHM and calculate its adaptive moments. 
We then evaluate the trace of the PSF-convolved galaxy image and of the PSF image from their moments using Eq.~\eqref{trace}, and evaluate the ratio $T_{\rm PSF}/T_{\rm gal}$.
We also apply a resolution factor minimum cut, defined as $1 - \frac{T_{\rm PSF}}{T_{\rm PSF} + T_{\rm gal}}$ at $0.1$, to exclude galaxies that are too small to be resolved compared to stars. Finally, the list of trace ratios that passes the resolution factor cut is arithmetically averaged, giving a value of $2.10$ with a population standard deviation of 1.95 due to a long right-side tail of the distribution of ratios. This is close to that found by \citealt{jarvis} which was $2.42$. Given the broad distribution of galaxy sizes in real galaxy samples, the difference between these numbers represents a modest shift towards smaller galaxy size expected in LSST analysis compared to the DES analysis in \citealt{jarvis}.

\begin{figure*}
\centering
\subfloat[Radial model -- translational dithering: Y1]{\includegraphics[width=0.48\textwidth]{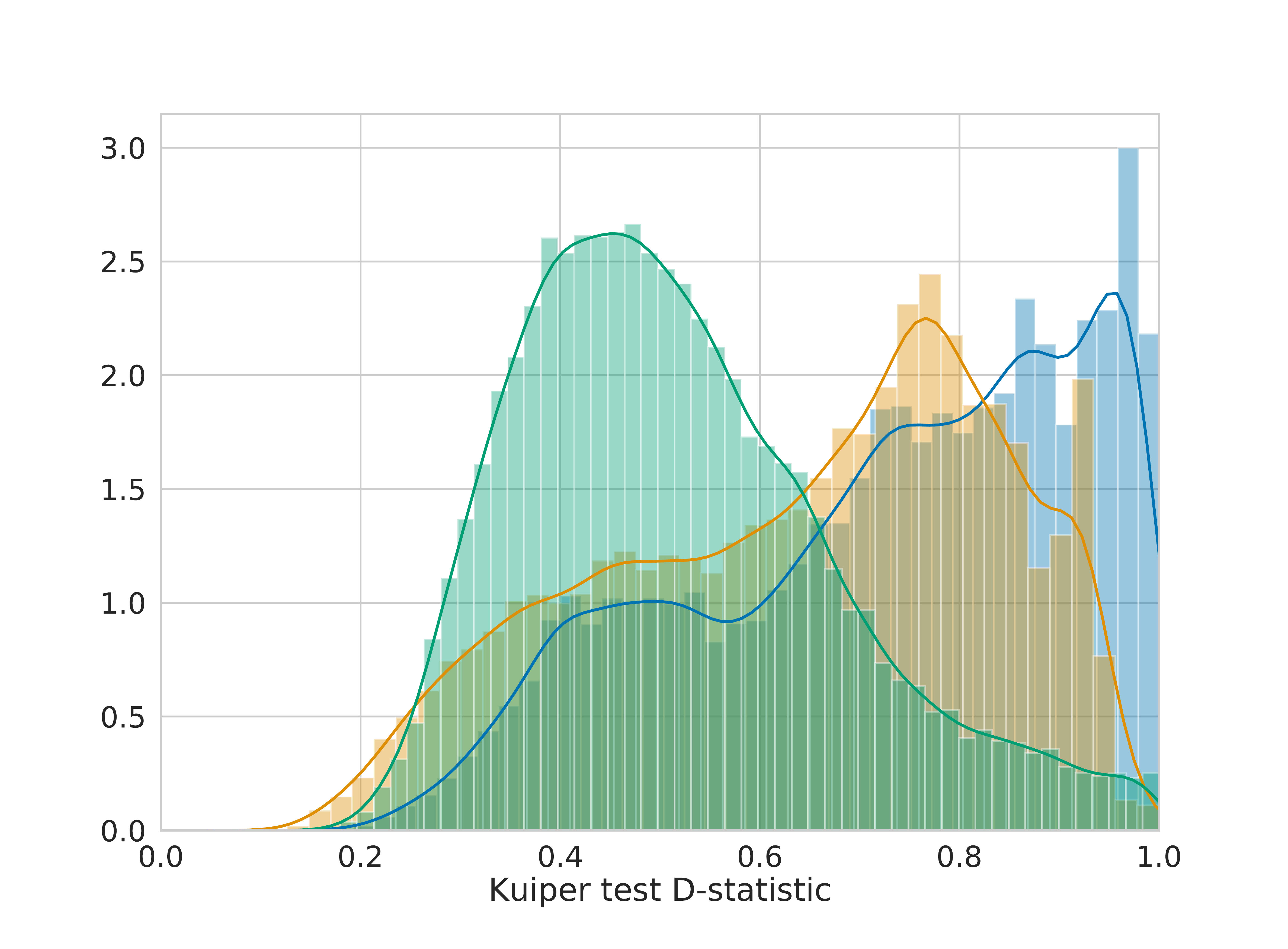}}\hfill
\subfloat[Radial model -- translational dithering: Y10] {\includegraphics[width=0.48\textwidth]{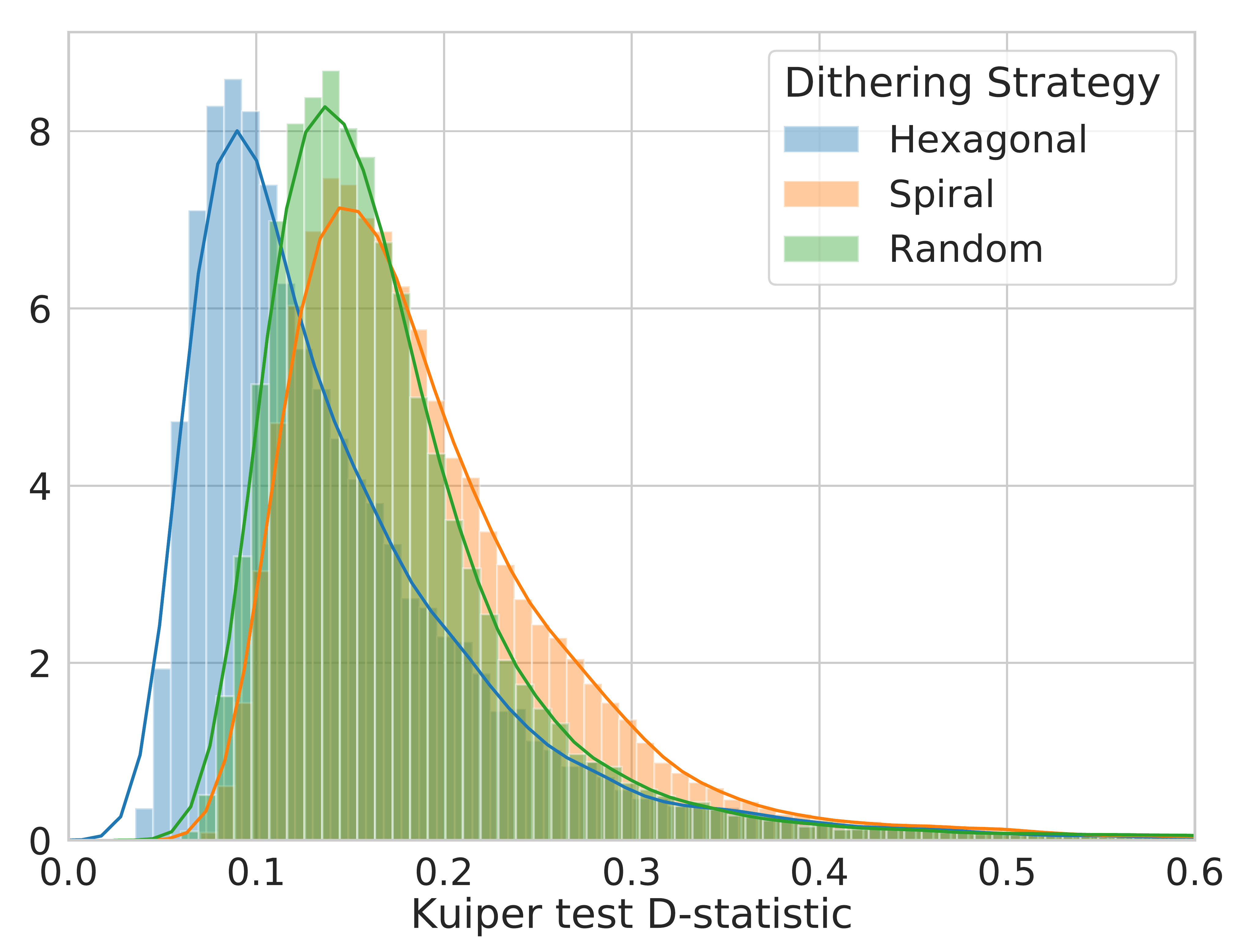}}\hfill

\subfloat[Horizontal model -- translational + rotational dithering: Y1]{\includegraphics[width=0.48\textwidth]{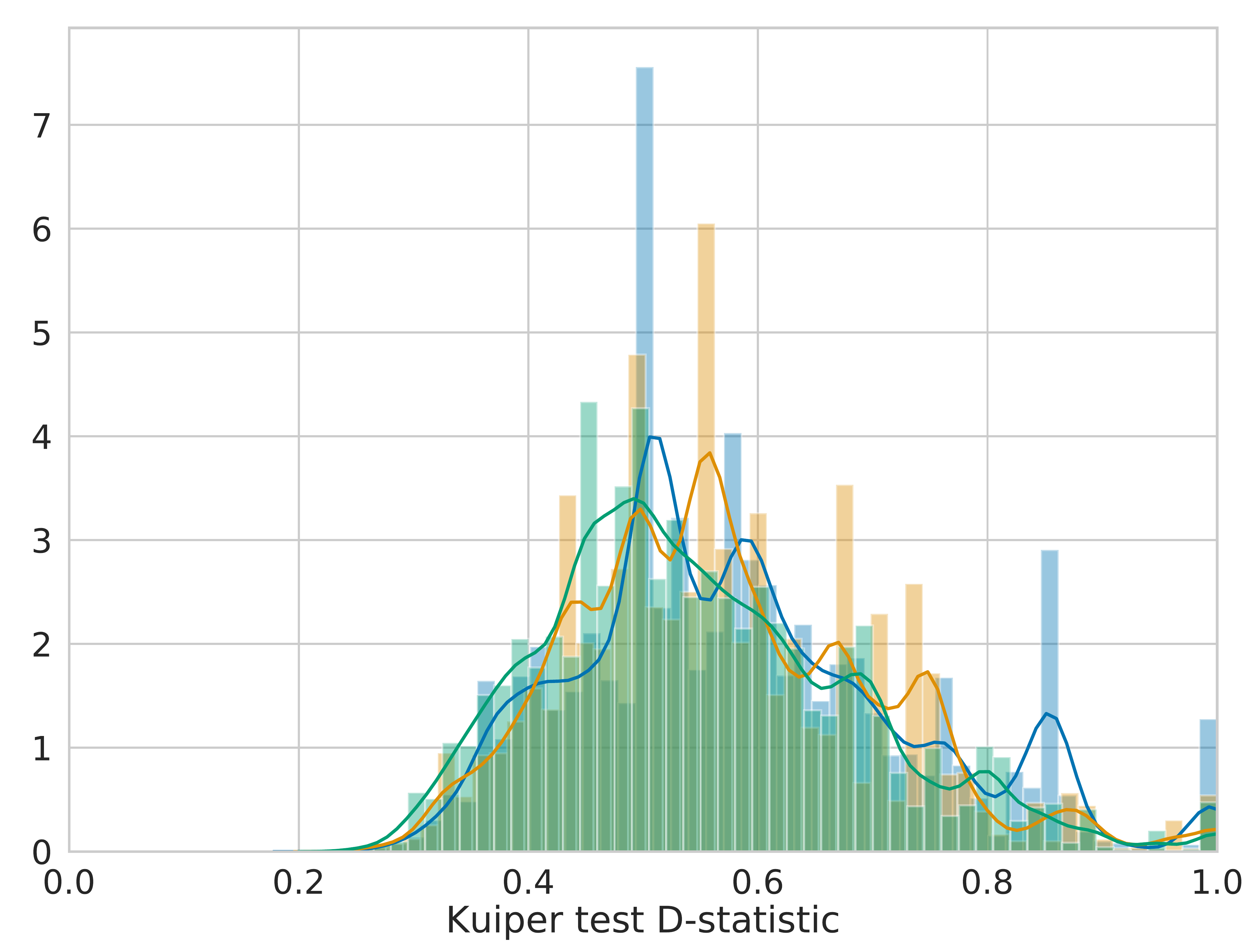}}\hfill
\subfloat[Horizontal model -- translational + rotational dithering: Y10] {\includegraphics[width=0.48\textwidth]{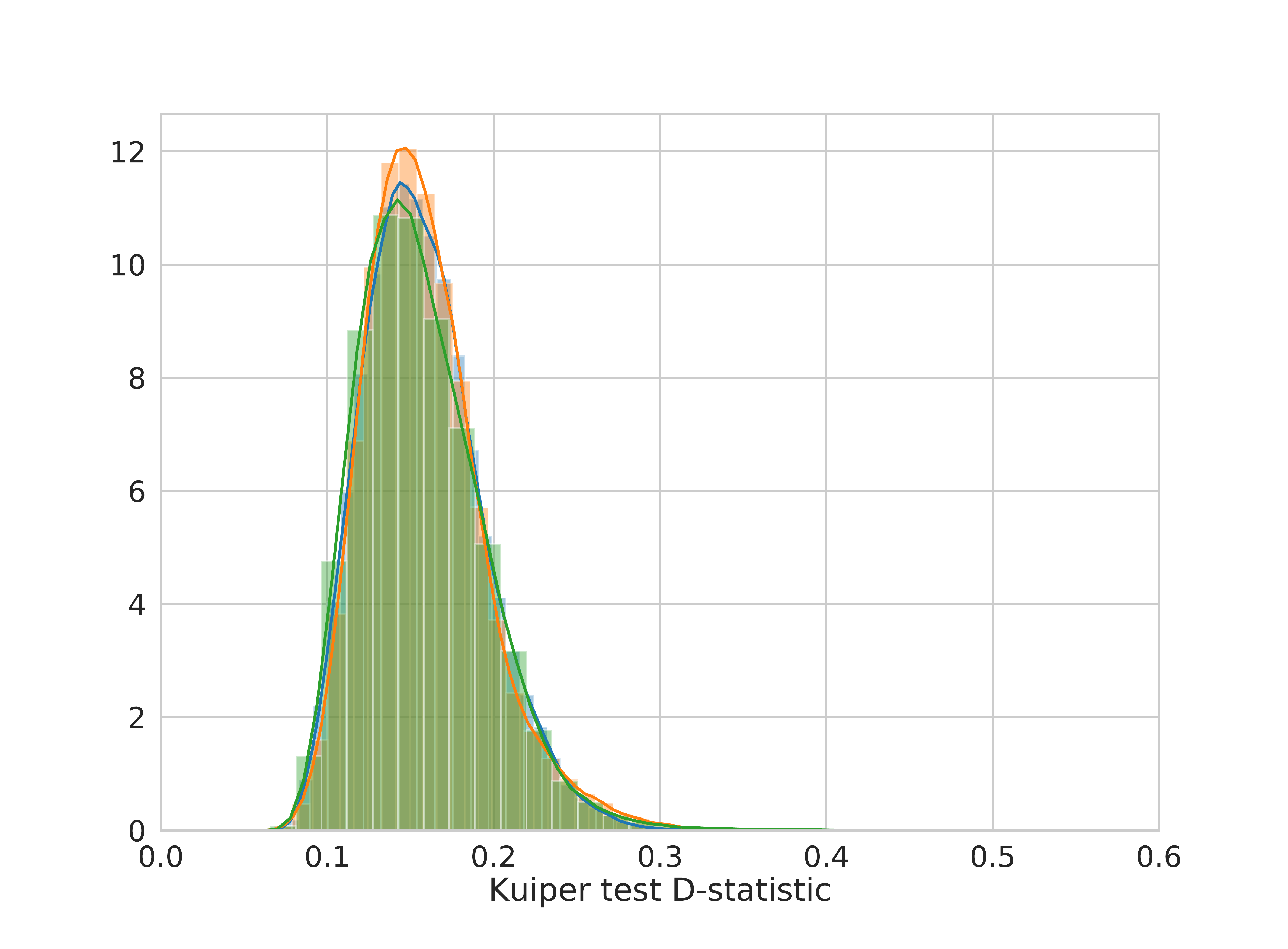}}\hfill

\caption{Distributions of the D-statistic from the Kuiper test for the four cases of the radial model with translational dithering applied at Y1 (a) and Y10 (b), and the horizontal model with both translational and rotational dithering applied at Y1 (c) and Y10 (d). Each value of the D-statistic comes from comparing the distribution of angles pointing from a single star to the dithered focal plane which observe it to a uniform distribution. For each of the three translational dithering strategies, the plots show the histograms and kernel density estimators of the D-statistic distributions. In (c), the distribution of D-statistics is not smooth due to two reasons: this distribution is made up of a small number of samples; and unlike the case in (a), the dithering timescale is once per filter change, which leads to a lot of the angle distributions looking almost the same, and the Kuiper test cannot distinguish one of them being more uniform than the other (especially that the multimodal behavior appears for D values larger than 0.5, where the Kuiper test is indicting that the distribution is very far from being uniform. At Y1, the random dithering strategy strongly outperforms the other strategies in the case of the radial model, and has average performance in the case of the horizontal model). At Y10, the random dithering strategy is outperformed by the hexagonal strategy in the case of the radial model, but the differences between all dithering strategies at Y10 are relatively small. An illustrative plot of two distributions contributing to the D-statistics shown here is presented in Fig.~\ref{fig:singleangledistribution}.
    } \label{fig:kstesttranslational}
\end{figure*}

\begin{figure*}
    \centering
    \includegraphics[width=0.48\textwidth]{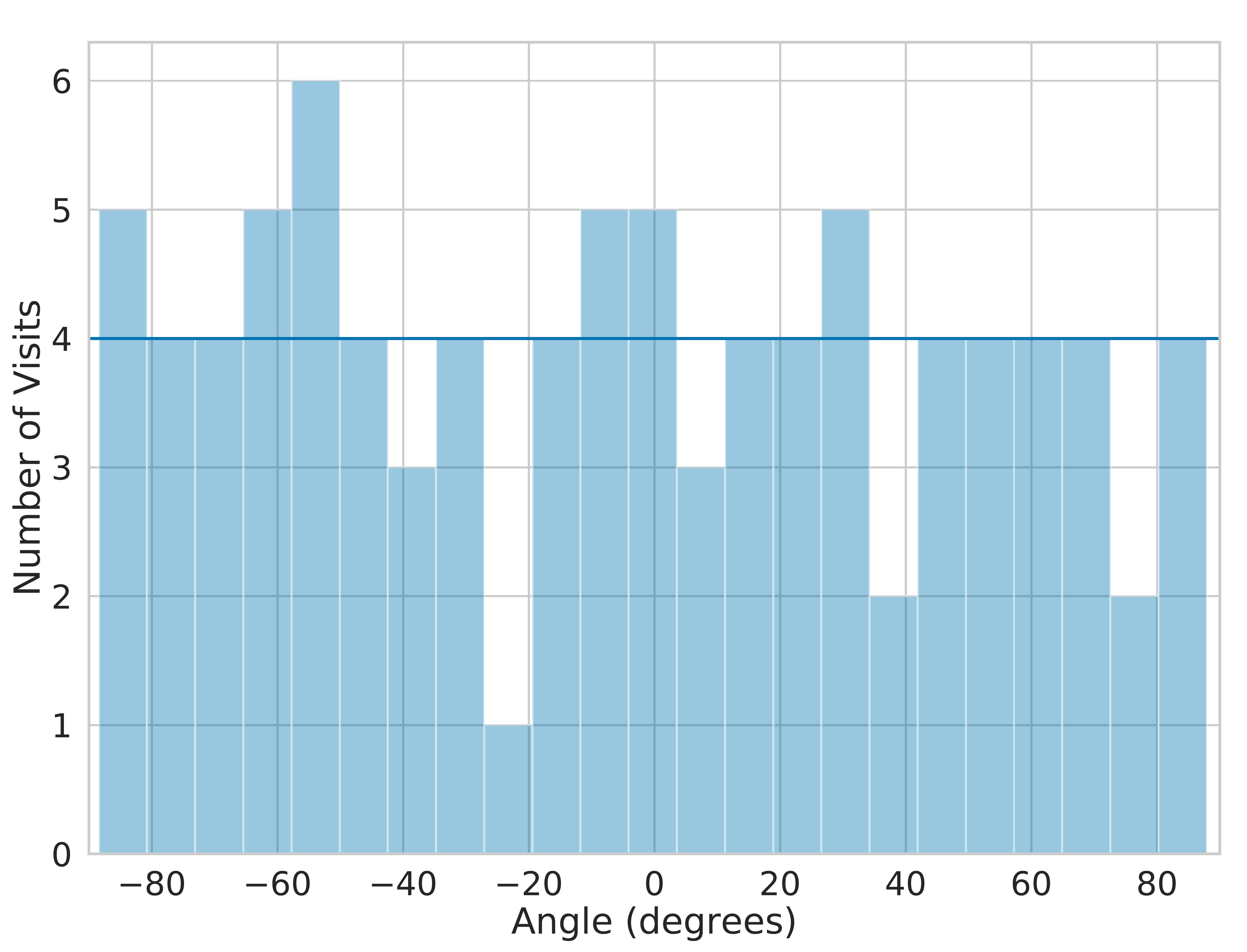}
    \includegraphics[width=0.48\textwidth]{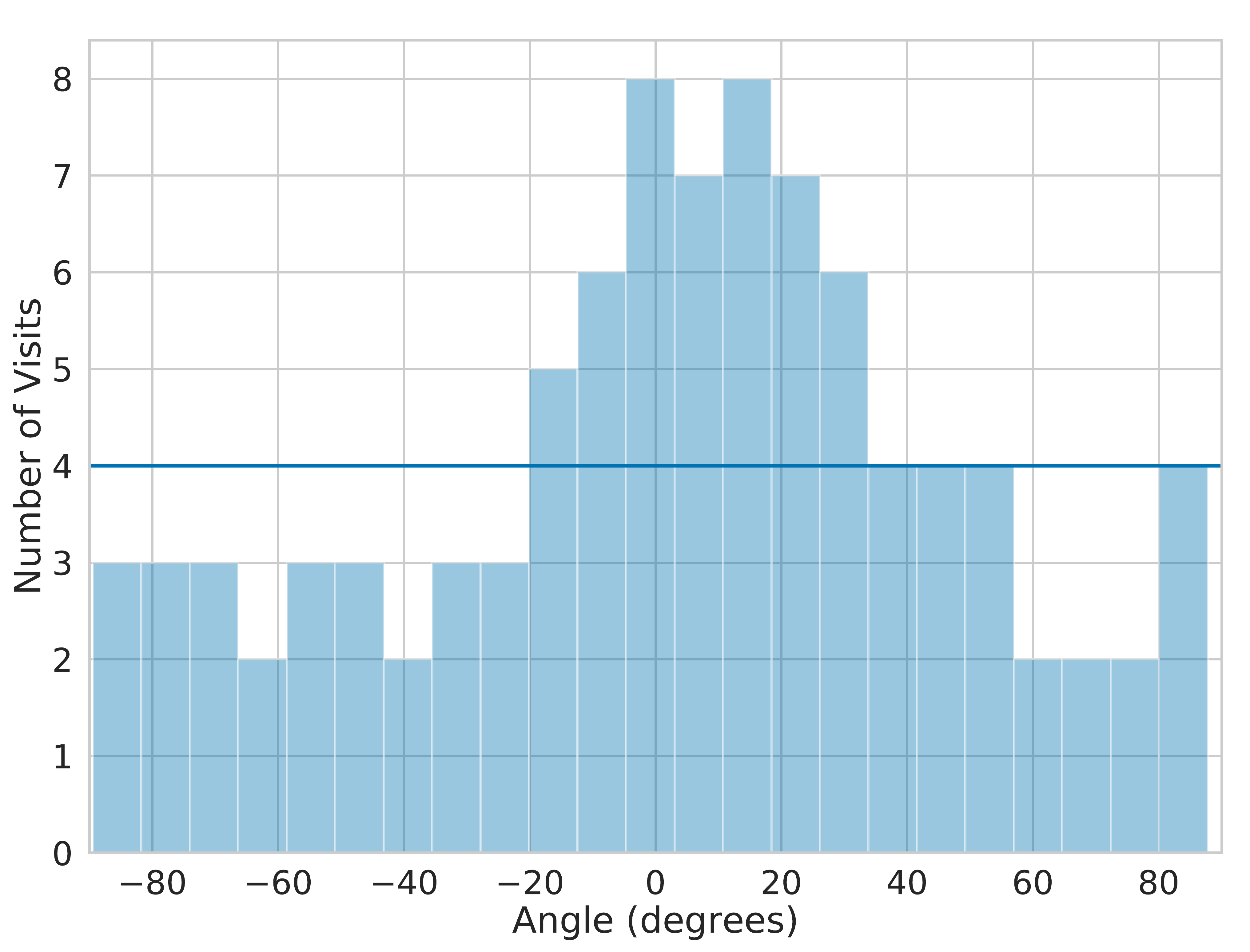}
    \caption{Illustration of two of the distributions that contribute to our evaluations of the D-statistics. These distributions are of the angles between +x and the lines from centres of the focal plane to an observed star. These plots correspond to the case of the hexagonal dithering at Y10, applied on the radial model. Left: 5th percentile (corresponding to a D-statistic of 0.06), right: 95th percentile (corresponding to a D-statistic of 0.23). A line corresponding to a uniform distribution has also been overplotted. It can be seen that the plot on the left is much closer to a uniform distribution than the one on the right.}
    \label{fig:singleangledistribution}
\end{figure*}

\section{Results}
\label{results}

First, we present results from the comparison of different dithering strategies applied to the baseline strategy, \texttt{baseline2018a}. We then choose the best dithering strategy and use it for the rest of this section as we explore the importance of other observing strategy choices, such as varying exposure time and area coverage. In studying dithering strategies, we start with a statistical uniformity test, and then present the full analysis (i.e., the effect on cosmic shear). When studying other observing strategy choices, we present the full analysis, then describe a simple proxy metric that provides a consistent estimation of the performance rankings of different strategies. We have made this metric available via MAF.

\subsection{Dithering Strategies}
\label{sec:ditherstrategies}
Previous metrics in \cite{COSEP} discussed in Section~\ref{sec:maf} do not show significant differences between different dither patterns, but rather indicate that rotational dithering is helpful in beating down systematics related to the parallactic angle. 

In this subsection, we consider statistical tests as well as the bias induced in the cosmic shear signal due to the discussed systematics, contrasting different combinations of translational and rotational dithering, extending the previous work to models of other additive WL systematics.

In all cases, we study the dithering on a per-visit timescale. Given that on average the LSST is designed to return to the same field position twice per night, choosing a per-night dithering strategy will, on average, multiply the additive bias on the cosmic shear by 2. Choosing a different timescale will, in general, multiply the additive bias on the cosmic shear by a constant factor, without affecting the relative rankings between the strategies we consider.

\subsubsection{Statistical Tests of Uniformity}
\label{sec:stattests}
Fig~\ref{fig:kstesttranslational} shows the values of the D-statistic from the Kuiper test described in Section~\ref{psfmodelingerrors}, computed using Astropy \citep{astropy} for 100,000 stars in the (RA, Dec) field described in Section~\ref{method}. The sample distributions used in the Kuiper test come from each star, where we compile a distribution of angles between the lines from the stars to all the centres of focal plane that can observe this star and the +x axis. The result is presented for all dithering strategies, for both the radial and horizontal toy models for systematics, and at both Y1 and Y10. D-statistics closer to 0 correspond to the distribution of angles being closer to uniform. Given that these angles are used to define coordinate systems for galaxy shapes, and that ellipticities are spin-2 quantities, the results are shown modulo 180 degrees. We use angle distributions rather than ellipticities here due to their interpretability. It is important to note, however, that given our simple models, there is a one-to-one mapping between these two quantities. At Y1, random dithering provides the best systematics mitigation out of the three dithering strategies considered here, particularly in the case of the radial model. At Y10, random dithering has moderate performance compared to hexagonal and spiral dithering. This Kuiper test is less discriminating in later years (i.e., Y10), where the D-statistic values have shifted closer to 0 compared to Y1. It is also less discriminating than a full analysis that computes the additive bias on the cosmic shear, since the latter involves computing the shear bias to the second power. The Kuiper test does, however, preserve the ranking of the performance of the dithering patterns. Fig.~\ref{fig:singleangledistribution} shows illustrative plots of two of the angle distributions (specifically, in the case of hexagonal dithering applied to the radial model at Y10). The figure compares distributions at the 5th percentile and 95th percentile of the D-statistics values, to illustrate their difference in uniformity.

\subsubsection{Bias Induced in Cosmic Shear}

\begin{figure*}
    \centering
        \subfloat[Radial model -- translational dithering: Y 1]{\includegraphics[width=0.49\textwidth]{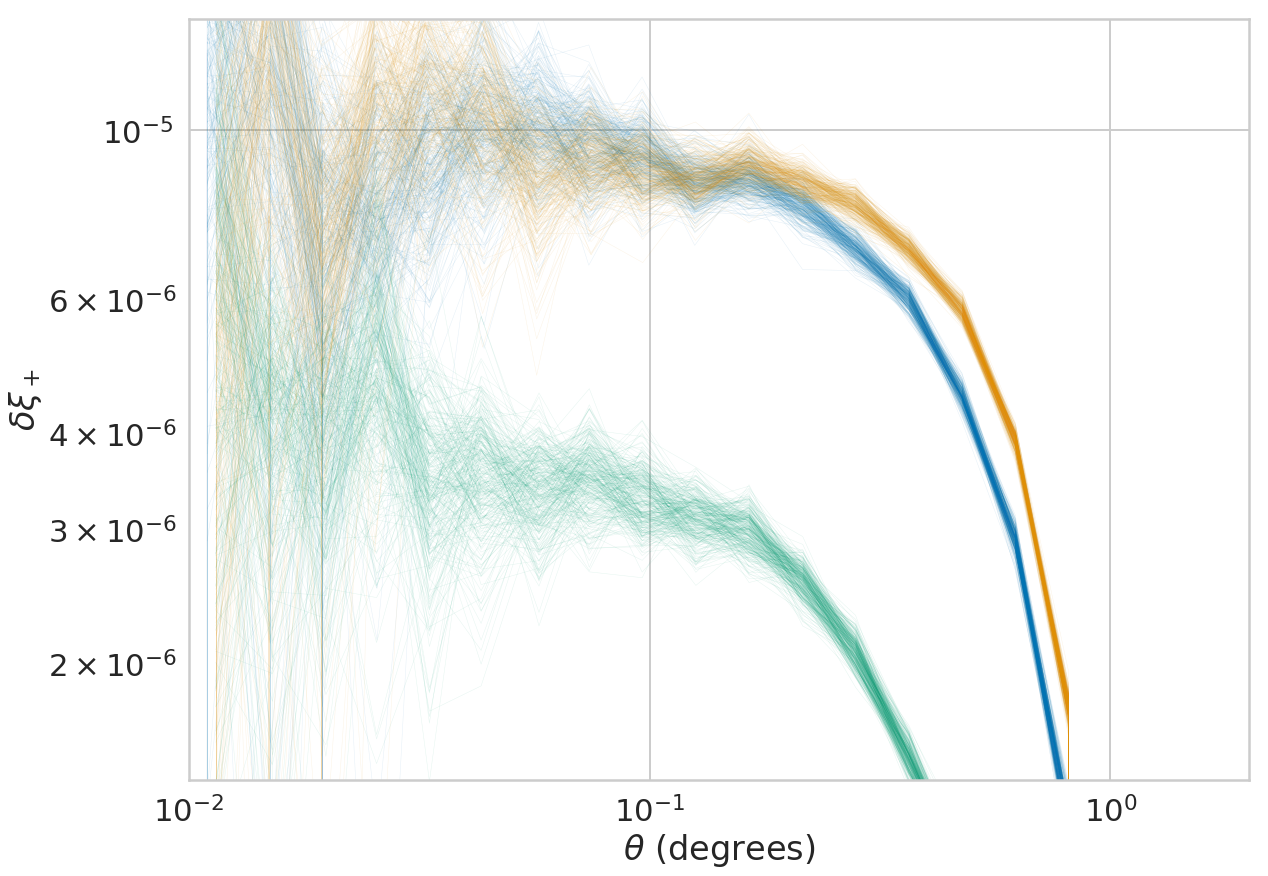}}\hfill
        \subfloat[Radial model -- translational dithering: Y 10] {\includegraphics[width=0.49\textwidth]{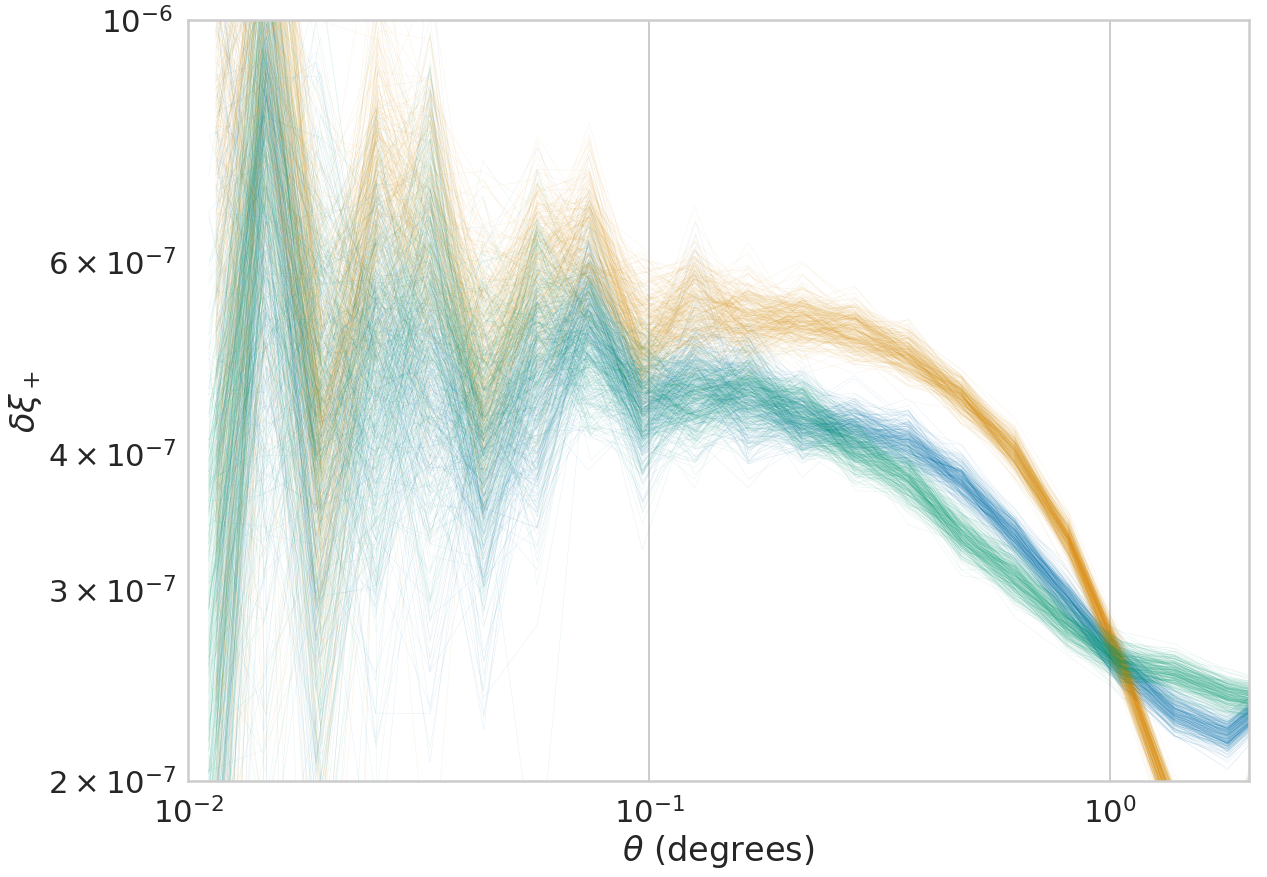}} \hfill
        \subfloat[Horizontal model -- translational + rotational dithering: Y 1]{\includegraphics[width=0.5\textwidth]{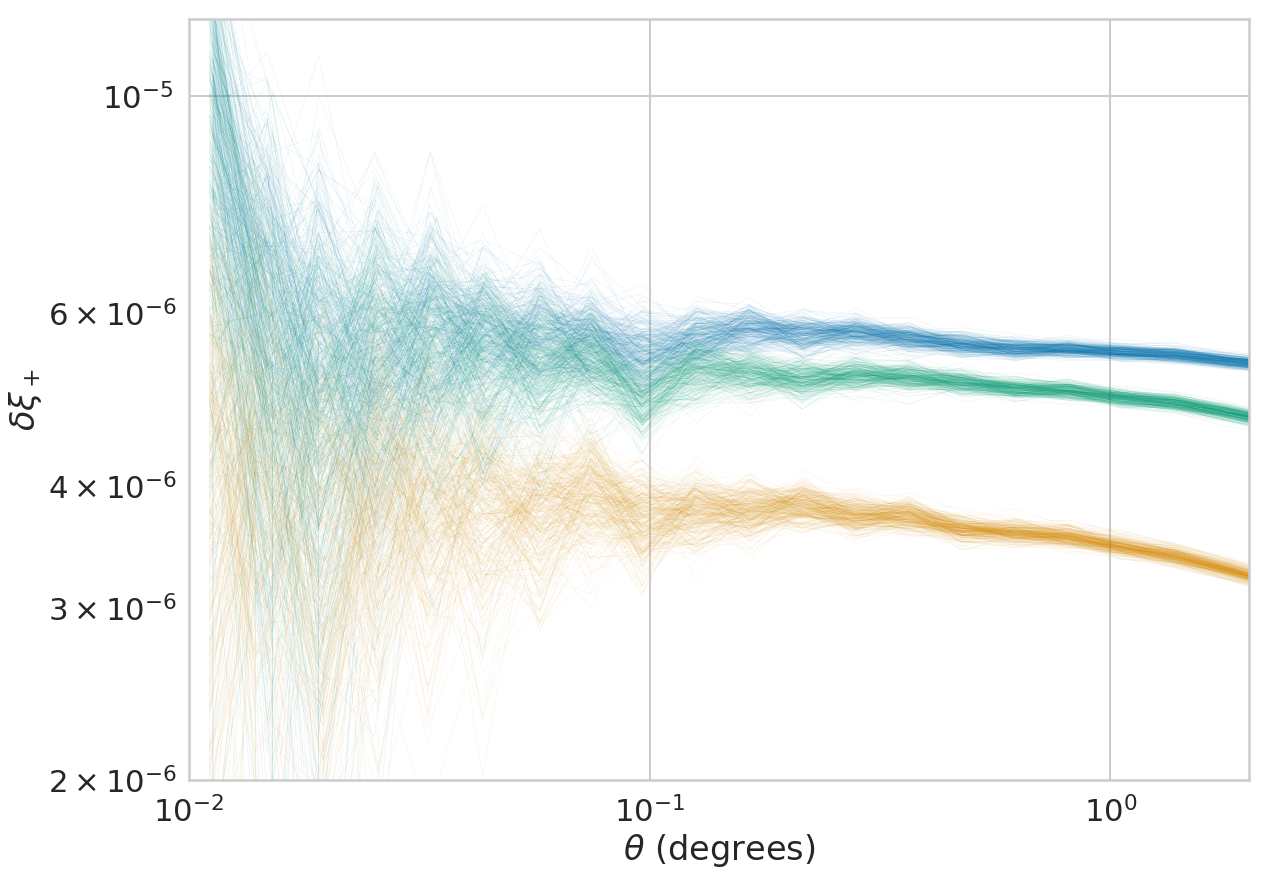}}\hfill
        \subfloat[Horizontal model -- translational + rotational dithering: Y 10] {\includegraphics[width=0.5\textwidth]{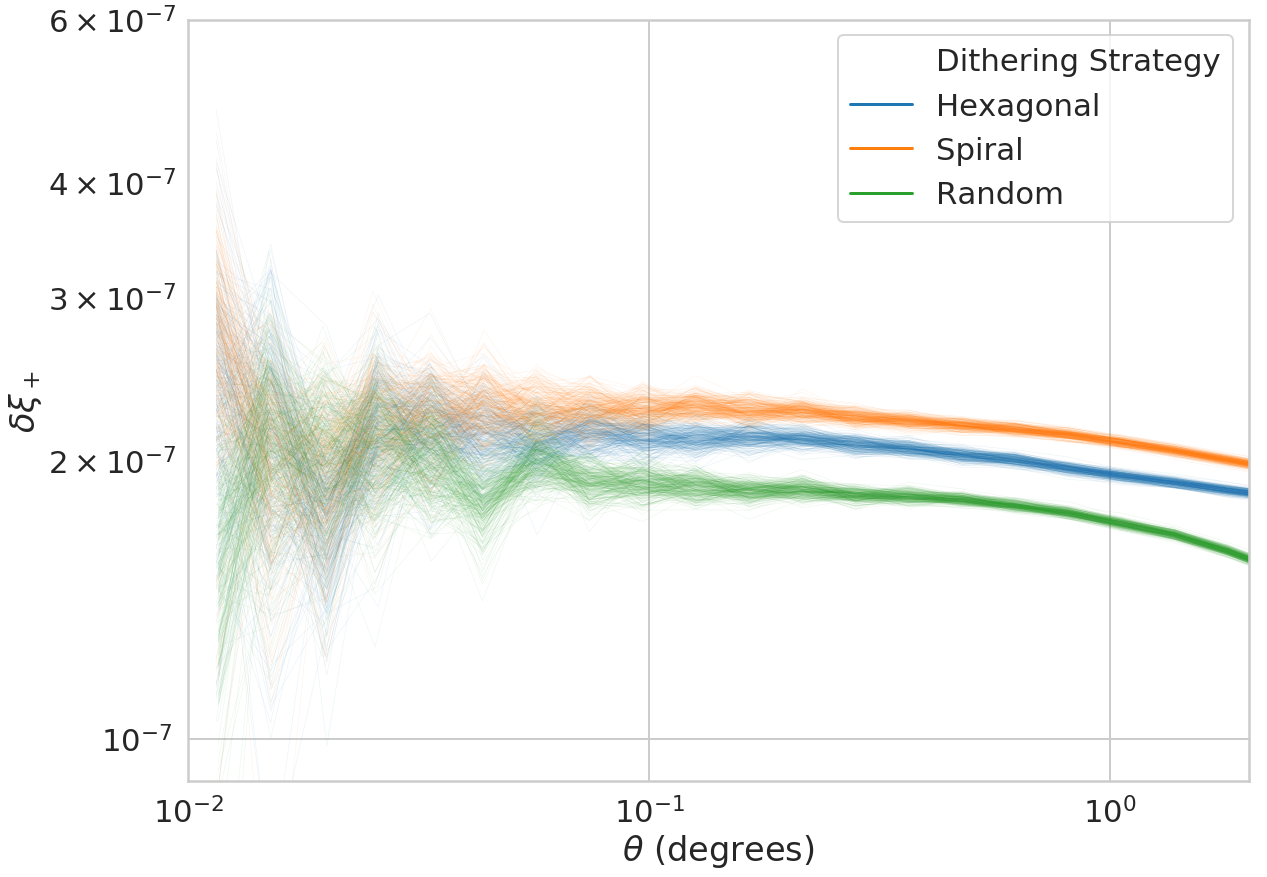}}\hfill
    \caption{Comparison between the additive systematic bias in the cosmic shear signal after propagating the PSF model residuals using the radial and horizontal models as indicated, and the formalism in Section~\protect\ref{cosmoprop}, for the three  translational dithering strategies described in Fig.~\ref{fig:ditherpatterns}. These dithering strategies are applied to the \texttt{baseline2018a} reference OpSim run. The plots provide a ranking of the patterns for Y1 and Y10; the random dithering strategy outperforms the other options except for the case of the horizontal model at Y1.    The plots show 300 realizations obtained by bootstrapping the results for the stars used to calculate the correlation functions, to show the scatter. Note that the vertical axes span different ranges in different panels, and in particular, for example, (a) shows that random strategy mitigates weak lensing additive systematics by a factor of 2--3, while in (d) the differences across all strategies tiny. 
    \label{fig:ditherresults}
    }
\end{figure*}

To compute the bias in the cosmic shear, we use the formalism in Section~\ref{cosmoprop}, using  TreeCorr (\citealt{treecorr2}, \citealt{treecorr}) to compute the correlation functions between 0.01 and 10 degrees in 26 logarithmically spaced bins. 
Fig.~\ref{fig:ditherresults} shows the additive bias in the cosmic shear after Y1 and Y10 for the three translational dithering strategies applied to \texttt{baseline2018a}. The absolute magnitude of the curves depends on the specific numbers in our toy models, and thus only the relative ordering of the curves is meaningful.  These plots are consistent with the results from the simpler statistical tests in Fig.~\ref{fig:kstesttranslational}.

Random dithering is the best-performing dither pattern for all cases except for the horizontal model at Y1. \citet{humna} also found that random dithering leads to the best performance when quantifying the effect of dithering strategies on large-scale structure systematics. For these reasons, we choose the random translational dithering strategy in the following subsections.

\begin{figure*}
    \centering
        \subfloat[Year 1]{\includegraphics[width=0.49\textwidth]{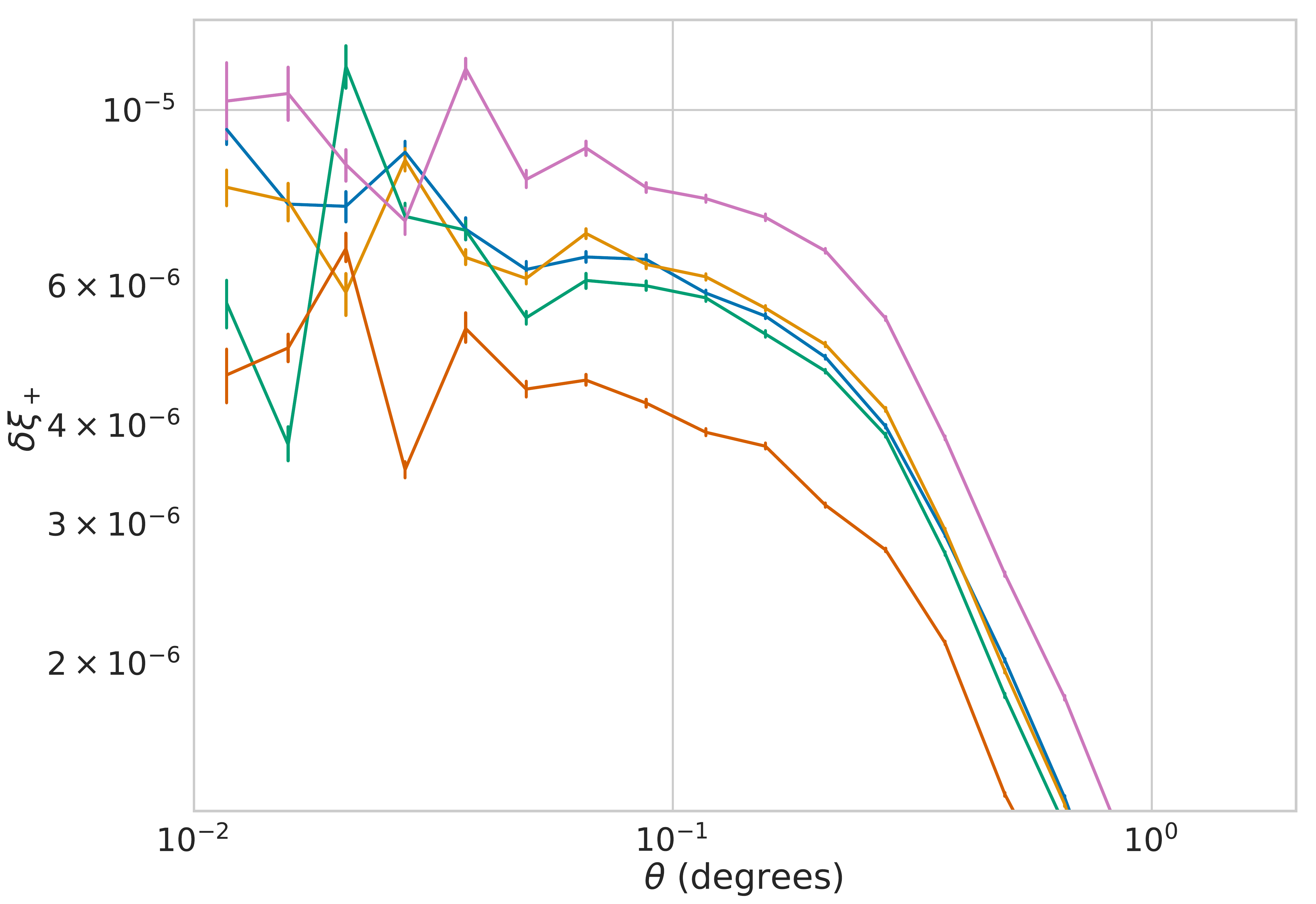}}
        \subfloat[Year 10] {\includegraphics[width=0.49\textwidth]{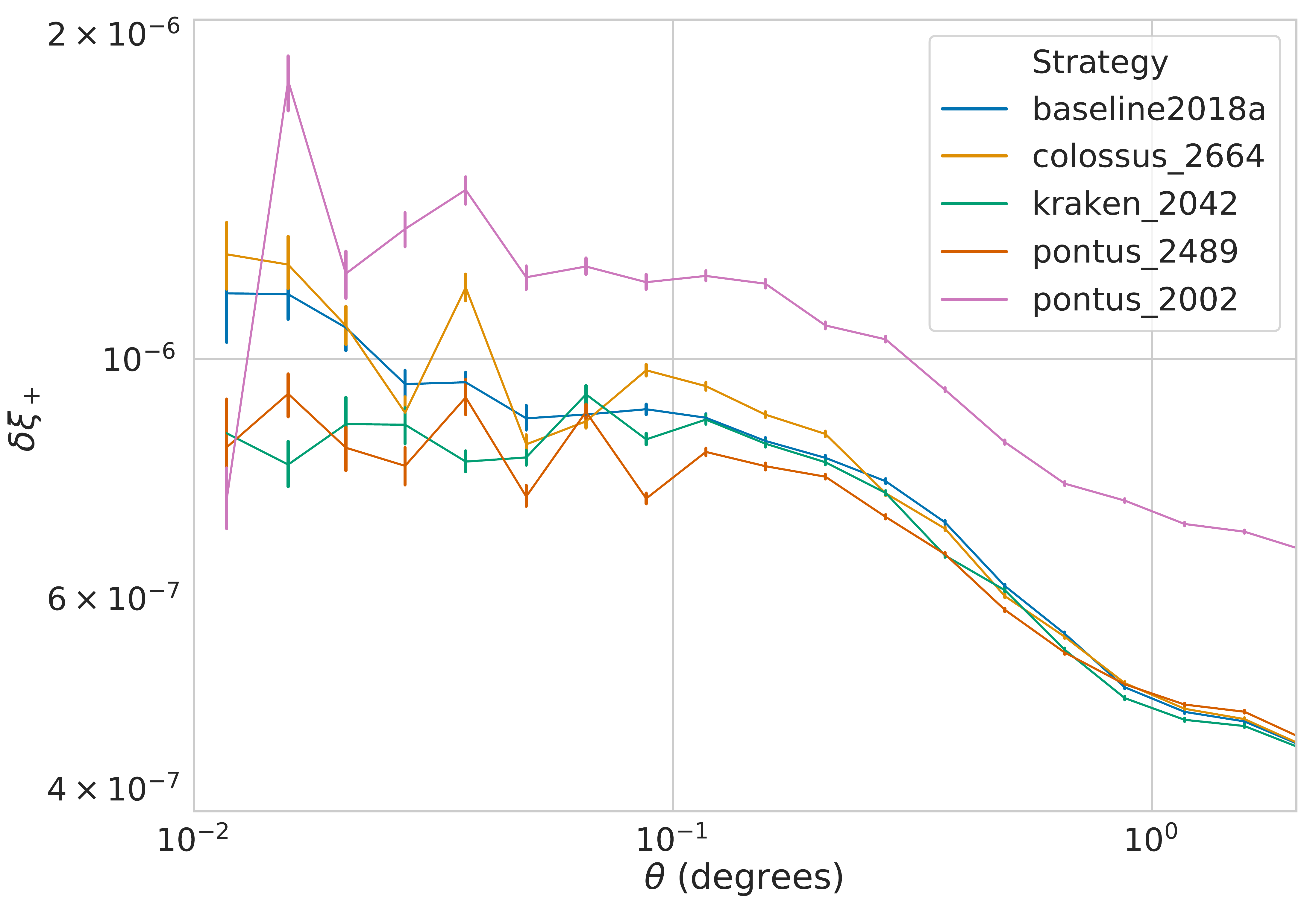}} 
        \caption{A comparison of the additive systematic biases in the cosmic shear signal for the strategies in Table~\ref{tab:runs} at Y1 and Y10 of the survey. These biases were obtained using the systematic error propagation formula in Section~\protect\ref{cosmoprop}, for the toy model with radial residuals from Section~\protect\ref{method}. The relative magnitudes of these curves provides a meaningful ranking of the strategies, with lower systematic bias being preferred, while the absolute magnitude is arbitrary.  Note that the scales spanned by the vertical axes on the two panels are different. } \label{fig:stratcomp}
\end{figure*}

\subsection{Other Aspects of Observing Strategy}

We now focus our attention on how other aspects of observing strategy affect weak lensing additive systematics, adopting random dithering applied to all OpSim runs that we consider in this work, summarised in Table~\ref{tab:runs}. 

\subsubsection{Effect on Cosmic Shear}
Fig.~\ref{fig:stratcomp} shows the additive bias on the cosmic shear signal for the survey strategies studied. As mentioned before, only the relative ranking of strategies is meaningful. The plot shows that the larger-area strategy (\texttt{pontus\_2002}) performs much worse than the baseline; strategies that spend more time in the Galactic plane (\texttt{colossus\_2664}) perform slightly worse; the strategy with 30-second single visits (\texttt{kraken\_2042}) does slightly better than the baseline 2$\times$15s; and the strategy with 20-second single visits \texttt{pontus\_2489} performs significantly better than the baseline. Physical reasoning for these results is provided below. It would be expected for a rolling cadence strategy that begins rolling during the first year of the survey to perform better than \texttt{baseline2018a} during Y1 due to its focusing on a smaller area within this year, and to perform similarly well to \texttt{baseline2018a} at Y10 assuming that the rolling cadence strategy is gently rolling and assures similar survey uniformity across the footprint by the end of the survey.

\subsubsection[Counts Metric]{Counts Metric\footnote{\url{https://github.com/lsst/sims_maf/blob/master/python/lsst/sims/maf/metrics/weakLensingSystematicsMetric.py}}}
\label{sec:metric}
\begin{figure*}
    \centering
        \includegraphics[width=0.85\textwidth]{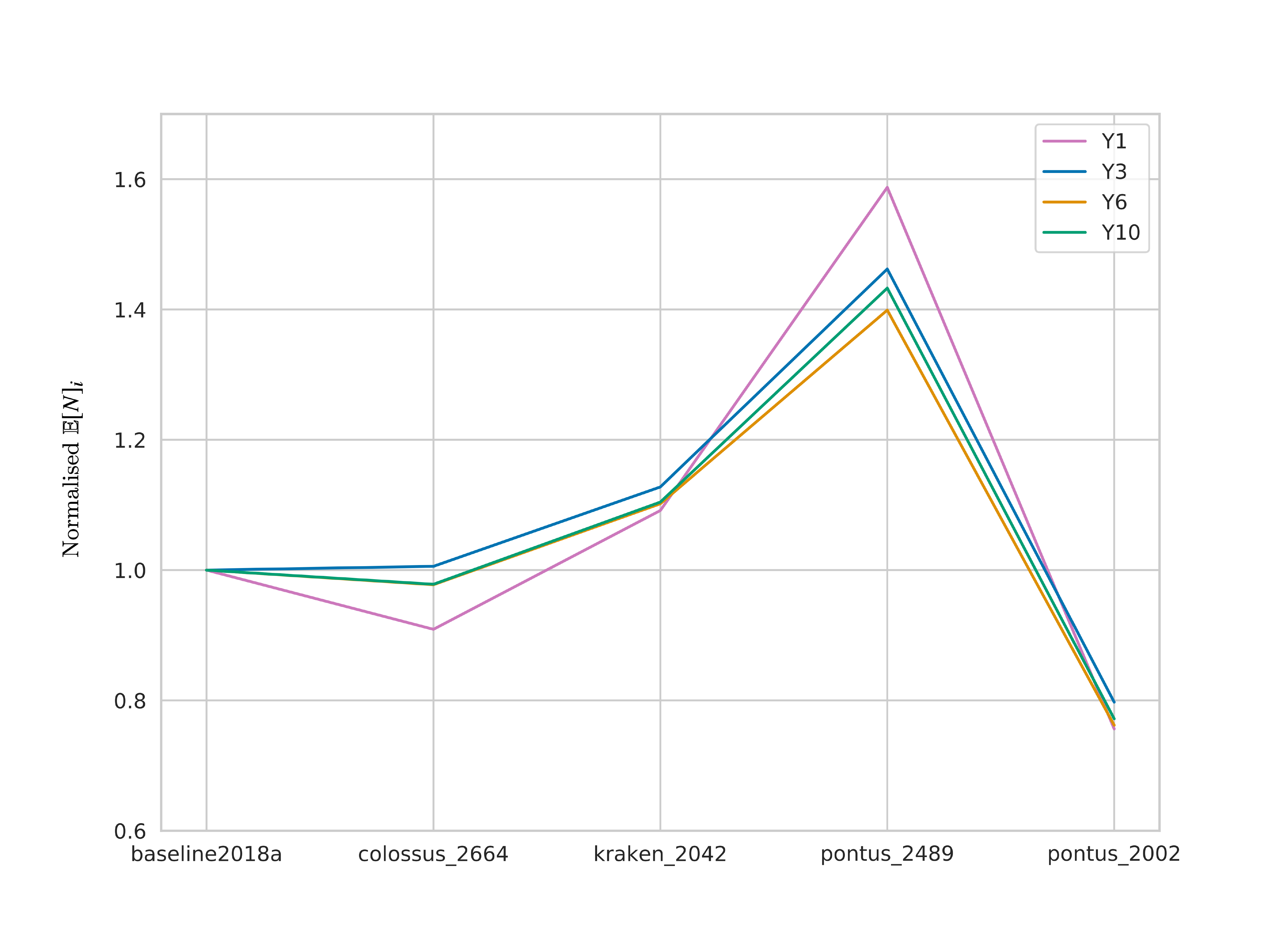}
        \caption{The average number of $i$-band exposures for observing strategies described in Table~\ref{tab:runs} normalized to \texttt{baseline2018a} at each milestone. This metric provides a simple way to rank the performance of different observing strategy choices -- with higher number of exposures corresponding to better performance. The link between this metric and the error on cosmic shear is demonstrated in Table~\ref{tab:corr}. 
        }

        \label{fig:metric}
\end{figure*}

We again use a Kuiper test as a metric for ranking the observing strategies. However, since the dithering strategy we apply is the same for each OpSim run, only the number of observations will be different for the different strategies (and any other effects related to the distribution of angles pointing from the objects to the centres of focal planes will on average be the same for all the strategies). Therefore, we initially defined the metric here as the average number of observations for a set of objects randomly distributed in RA and Dec, observed in the $i$-band. We further developed this metric to be easily run within MAF, and therefore some development choices, such as replacing the 100,000 random positions with a sparse HEALPix\footnote{\url{https://healpix.sourceforge.io}} \citep{healpix} grid, have been adopted. Empirical checks show that a HEALPix grid with as few as 5,000 cells (equivalent to a HEALPix Nside specification of 32) yields consistent results with 100,000 randomly-sampled objects from a uniform distribution. While this is sufficient for the counts metric, a HEALPix grid of 5,000 cells is not sufficient for the full correlation function-based analysis because (a) the number of objects is not sufficient to measure precise small-scale correlations, and (b) gridded input data, when used as an input to tree-based correlation function estimators such as TreeCorr, may induce spurious features in the correlation function. 
The counts metric is plotted as a function of observing strategy in Fig.~\ref{fig:metric}, which provides consistent results with Fig.~\ref{fig:stratcomp}. This metric also explains why some strategies perform better: additive WL systematics, such as the ones studied here, average down with the number of exposures, since more exposures lead to the distribution of angles pointing towards the centre of the focal plane becoming more uniform, given that the same dithering strategy is adopted. \texttt{pontus\_2002} covers a larger area in the same amount of time, so each object is observed fewer times on average. \texttt{colossus\_2664} spends more time in the Galactic plane, which is not used in the weak lensing analysis due to the high extinction, reducing the time available for the WFD survey in the areas that pass our cut and again lowering the average number of observations for each object. \texttt{kraken\_2042} makes single 30-second observations rather than two 15-second observations, eliminating the read-out time in between, allowing for more time to observe the same area, and thus providing a larger number of observations. \texttt{pontus\_2489} makes single 20-second observations in most bands, which allows for even more observations. 

Table \ref{tab:corr} shows the relative magnitudes of the cosmic shear bias normalised to the \texttt{baseline2018a} strategy, based on a $\chi^2$ fit; as well as the correlation coefficient between the metric values and the $\chi^2$ fits. We see strong (negative) correlation at Y1, and moderate (negative) correlation at Y10. The reason is that at Y10, once $\mathbb{E} [N]_i \sim 225$ visits, the distribution of angles between the line connecting the star locations to focal plane centres and +x is sufficiently sampled, making for a rotationally uniform distribution. In conclusion, this simple proxy metric can be used at Y1 to clearly rank the different strategies, while it can be used in Y10 to detect particularly bad strategies for WL systematics, such as \texttt{pontus\_2002} (although this strategy does not meet LSST SRD requirements for median number of visits).

The weak lensing analysis in general will use multiple bands, not just $i$-band; the choice of bands used is driven by PSF modelling adequacy and signal-to-noise ratio considerations, and will most likely include $r$, $i$, and $z$.  Studies of strategies that have a different distribution of time spent observing in each filter (e.g., for strategies that spend more time on $i$ vs. $r$, or vice versa) should take into account that while the metric allows for using different filters or combinations thereof, results from different filters may not be directly comparable with the metric results in this paper. The metric will also not be robust if different strategies use different dithering algorithms -- in those cases, using a Kuiper test in the way described in Section~\ref{sec:stattests} would be necessary for a fair comparison.

\begin{table*}
    \centering
\begin{tabular}{c|cccccc}

    Year  & baseline2018a &  c\_2664 & k\_2042 & p\_2489 & p\_2002 & Pearson-r \\\hline\hline
    1  & 1 & 1.00 & 0.96 & 0.82 & 1.42 & -0.84 \\
    10  & 1 & 1.00 & 0.99 & 0.99 & 1.46 & -0.64 \\
\end{tabular}
    \caption{the relative magnitude of the cosmic shear bias normalised to the \texttt{baseline2018a} strategy, based on a $\chi^2$ fit; lower numbers correspond to better performance. To demonstrate the usefulness of our proxy metric, the Pearson r-correlation coefficient is also reported  between the proxy metric values and the best-fit numbers in the table. This relation is stronger at Y1, while it eventually gets saturated at Y10 for runs with more than 230 average $i$-band visits, when the observed distribution of the pointing from objects to focal plane centres gets sufficiently sampled}
    \label{tab:corr}
\end{table*}


\section{Conclusion}
\label{sec:conclusion}

The LSST will provide new opportunities for weak lensing and dark energy science in general. The LSST observing strategy will affect these science cases in different ways.  Thus, exploring the impacts for individual science cases is essential to optimising the observing strategy. The LSST also provides new opportunities for systematics mitigation due to its unique dithering (in scale and number) and observing strategies.

We used models of additive cosmic shear systematics and simulated how they are affected by dithering and other observing strategy considerations, such as variations in area and exposure time, using several LSST operational simulations. Using a formalism to propagate these models into additive shear bias, as well as simpler metrics, we conclude that additive cosmic shear systematics will average down best with (a) random translational dithering (applied in addition to random rotational dithering at every filter change), and (b) with higher numbers of visits in the WFD survey area. These results are not necessarily the same for other science cases (or even for WL statistical constraining power), and will eventually be used in conjunction with how the observing strategy affects other science cases to recommend an optimal observing strategy for the LSST.

The study in this paper has only considered WL cosmic shear systematics. There is a depth-area tradeoff between cosmic shear systematics and WL statistical constraining power, where it was found in \cite{OSTFWhitepaper} that the WL constraining power favors survey strategies with larger areas, since the change in area has a larger effect than the loss of the average number of visits (and consequently, assuming the same visit duration, the loss in depth). A full exploration of the tradeoff between the impact of observing strategy on the statistical constraining power for cosmology versus for systematics mitigation is an important part of future work.

\section*{Contributors}

HA: Lead and corresponding author, worked on statistical  and computational formal analysis, research investigation,  methodology development, software development, visualization, and writing.
RM: Scientific oversight, paper editing.
HAw: Provided feedback on dithering details, alongside comments on draft.
JM: Provided Fig.~4 and scientific suggestions and comments on draft.
JAT: Wrote some text on systematics, and optimal angle dithering to suppress systematics.
PY: Support for writing MAF metrics and understanding simulated survey databases.
EG: Provided scientific suggestions and comments on draft.

\subsection*{Acknowledgments}

HA and RM are supported by the Department of Energy Cosmic Frontier program, grant DE-SC0010118.

HA \& HAw 
thank the LSSTC Data Science Fellowship Program, which is funded by LSSTC, NSF Cybertraining Grant \#1829740, the Brinson Foundation, and the Moore Foundation; their
participation in the program has benefited this work.
  
HAw has been supported by the Rutgers Discovery Informatics Institute (RDI2) Fellowship of Excellence in Computational and Data Science(AY 2017-2020) and Rutgers University \& Bevier Dissertation Completion Fellowship (AY 2019-2020). HAw and EG were supported by the Department of Energy Cosmic Frontier program, grants DE-SC00011636 \& DE-SC0010008. 

RLJ and PY acknowledge support from the DIRAC Institute in the Department of Astronomy at the University of Washington. The DIRAC Institute is supported through generous gifts from the Charles and Lisa Simonyi Fund for Arts and Sciences, and the Washington Research Foundation.

Part of this work was performed under the auspices of the U.S. Department of Energy by Lawrence Livermore National Laboratory under Contract DE-AC52-07NA27344.

JAT acknowledges support from Department of Energy grant DE-SC0009999 at UC Davis. 

Figures in this paper were generated using the open-source Python libraries: matplotlib \citep{matplotlib}, seaborn \citep{seaborn}.

The DESC acknowledges ongoing support from the Institut National de Physique Nucl\'eaire et de Physique des Particules in France; the Science \& Technology Facilities Council in the United Kingdom; and the Department of Energy, the National Science Foundation, and the LSST Corporation in the United States.  DESC uses resources of the IN2P3 Computing Center (CC-IN2P3--Lyon/Villeurbanne - France) funded by the Centre National de la Recherche Scientifique; the National Energy Research Scientific Computing Center, a DOE Office of Science User Facility supported by the Office of Science of the U.S.\ Department of Energy under Contract No.\ DE-AC02-05CH11231; STFC DiRAC HPC Facilities, funded by UK BIS National E-infrastructure capital grants; and the UK particle physics grid, supported by the GridPP Collaboration.  This work was performed in part under DOE Contract DE-AC02-76SF00515.

This paper has undergone internal review in the LSST Dark Energy Science Collaboration by Tim Eifler, Mike Jarvis, and Sukhdeep Singh, we thank them for their helpful comments and reviews.
We thank Seth Digel for constructive input on the paper.

\section*{Data Availability Statement}
The data underlying this article comprises of LSST Operational Simulations are available at \url{http://astro-lsst-01.astro.washington.edu:8080}; and data from the LSST Metrics Analysis Framework (MAF), available in Github at  \url{https://github.com/lsst/sims_maf} and can be accessed with unique commit identifier \texttt{34205f6c48}. Data associated with generating other plots in the article will be shared on reasonable request to the corresponding author.

\bibliographystyle{mnras}
\bibliography{main}

\end{document}